\documentclass{aa}  

\usepackage{graphicx}
\usepackage{txfonts}
\usepackage{graphicx}
\usepackage{placeins}
\usepackage{float}
\usepackage{subfigure}
\usepackage{amsmath}
\usepackage{mathrsfs}
\usepackage{amssymb}
\usepackage{color,epstopdf}
\usepackage{pdflscape}

\newcommand{\MRC}{MRC1138-262}
\newcommand{\Spider}{Spiderweb Galaxy}
\newcommand{\Lya}{Ly$\alpha$}
\newcommand{\Ha}{H$\alpha$}
\newcommand{\NII}{[N{\small II}]}
\newcommand{\SII}{[S{\small II}]}

\newcommand{\HeII}{He{\small II}}
\newcommand{\OII}{[O{\small II}]}
\newcommand{\OIII}{[O{\small III}]}
\newcommand{\CI}{[C{\small I}]{2--1}}
\newcommand{\CIn}{[C{\small I}]}
\newcommand{\CO}{CO(7--6)}
\newcommand{\Htwo}{H$_2$}
\newcommand{\water}{H$_2$O}
\newcommand{\dustcont}{246\,GHz}
\newcommand{\HzRG}{H$z$RG}

\newcommand{\HST}{\textit{HST}}

\newcommand{\compI}{component 1}
\newcommand{\compII}{component 2}

\begin{document} 

\title{ALMA Finds Dew Drops in the Dusty Spider's Web}

\titlerunning{Dew Drops in a Dusty Web}

   \author{Bitten Gullberg \inst{1,3,4}
    \and
        Matthew D. Lehnert \inst{2}
    \and
    Carlos De Breuck\inst{1}
    \and
    Steve Branchu\inst{1,5}
    \and
    Helmut Dannerbauer\inst{6}
    \and 
    Guillaume~Drouart\inst{7}
    \and
    Bjorn Emonts \inst{8}
    \and
    Pierre Guillard\inst{2}
    \and
    Nina Hatch\inst{9}
    \and
    Nicole P. H. Nesvadba \inst{10}
    \and
    Alain Omont\inst{2}
    \and
    Nick~Seymour\inst{11}
    \and
    Jo\"{e}l~Vernet\inst{1}
          }

   \institute{European Southern Observatory, 
              Karl-Schwarzschild-Str. 2, D-85748 Garching\\
              \email{bitten.gullberg@durham.ac.uk}
    \and
    Institut d'Astrophysique de Paris, UMR 7095, CNRS, Universit\'e Pierre et Marie Curie, 98bis boulevard Arago, 75014, Paris, France 
    \and
    Max-Planck-Institut f\"ur Extraterrestrische Physik, Giessenbachstra\ss e 1, 85748 Garching, Germany
    \and
    Centre for Extragalactic Astronomy, Department of Physics, Durham University,  South Road,  Durham DH1 3LE,  UK
    \and
    Universit\'e de Bordeaux, LAB, UMR 5804, 33270, Floirac, France
    \and
     Universit\"at Wien, Institut f\"ur Astrophysik, T\"urkenschanzstra\ss e 17, 1180, Wien, Austria
     \and
    Department of Earth and Space Science, Chalmers University of Technology, Onsala Space Observatory, 43992, Onsala, Sweden
    \and
    Centro de Astrobiolog\'ia (INTA-CSIC), Ctra de Torrej\'on a Ajalvir, km 4, 28850 Torrej\'n de Ardoz, Madrid, Spain
    \and
    School of Physics and Astronomy, University of Nottingham, University Park, Nottingham NG7 2RD, UK
    \and
    Institut d'Astrophysique Spatiale, CNRS, Universit\'e Paris-Sud, Bat. 120-121, F-91405 Orsay, France
    \and
    International Center for Radio Astronomy Research, Curtin University, GPO Box U1987, Perth, WA 6845, Australia
             }

   \date{Submitted October 26 2015, accepted February 10 2016}

 
\abstract{We present 0\farcs5 resolution ALMA detections of the
observed \dustcont\ continuum, \CIn\ $^3P_2\,\to\,^3$$P_1$
fine structure line (\CI), \CO\  and \water\ lines in
the $z=2.161$ radio galaxy \MRC, the ``\Spider''.  We detect strong
\CI\ emission both at the position of the radio core, and in a second
component $\sim$4\,kpc away from it. The 1100\,km/s broad \CI\
line in this latter component, combined with its H$_2$ mass of $1.6\times
10^{10}$\,M$_{\odot}$ implies this emission must come from a compact
region $<$60\,pc, possibly containing a second AGN. The combined H$_2$
mass derived for both objects using the \CI\ emission is
$3.3\times 10^{10}$\,M$_{\odot}$.  The total \CO/\CI\ line flux
ratio of 0.2 suggests a low excitation molecular gas reservoir and/or
enhanced atomic carbon in cosmic-ray dominated regions.

We detect spatially-resolved \water\ $2_{11}-2_{02}$ emission ---
for the first time in a high-$z$ un-lensed galaxy --- near the outer
radio lobe to the east, and near the bend of the radio jet to
the west of the radio galaxy.  No underlying \dustcont\
continuum emission is seen at either position. We suggest that the
\water\ emission is excited in the cooling region behind slow (10-40
km s$^{-1}$) shocks in dense molecular gas (10$^{3-5}$ cm$^{-3}$).
The extended water emission is likely evidence of the radio jet's impact
in cooling and forming molecules in the post-shocked gas
in the halo and inter-cluster gas similar to what is seen in low-$z$
clusters and other high-$z$ radio galaxies. These observations imply
that the passage of the radio jet in the interstellar and inter-cluster
medium not only heats gas to high temperatures as is commonly assumed or
found in simulations, but also induces cooling and dissipation
which can lead to substantial amounts of cold dense molecular gas. The
formation of molecules and strong dissipation in the halo gas
of \MRC\ may explain both the extended diffuse molecular gas and young
stars observed around \MRC .}


\maketitle
%

\section{Introduction}

The high-$z$ radio galaxy (\HzRG) \MRC\  at $z=2.161$ is one of the best
studied \HzRG\ \citep[e.g.][]{pentericci97, carilli97, pentericci98,
pentericci00, carilli02, stevens03, kurk04a, kurk04b, greve06, miley06,
hatch08, humphrey08, hatch09, kuiper11, ogle12, seymour12}, with a well
sampled spectral energy distribution (SED) covering from radio to X-ray.
Its radio morphology is typical of distant radio galaxies with a string
of radio bright knots along the radio jet extending to the west and a
single lobe to the east of the central radio core \citep{carilli97}.
The radio core has an ``ultra steep'' spectral index of $\alpha=-1.2$,
and the spectral indices of the radio knots systematically
steepen with increasing distance from the core \citep{pentericci97}.

The radio source is embedded in an environment over-dense in galaxies
over scales of hundreds of kpc \citep{pentericci98, miley06} to Mpc
scales \citep{pentericci02, kurk04a,kurk04b, dannerbauer14}.  Many of
these galaxies are clumpy and star forming \citep{pentericci98,
miley06, dannerbauer14}. \cite{hatch09} predict that most of the
satellite galaxies within 150\,kpc will merge with the central \HzRG\
and that the final merger galaxy will contain very little gas due to
the high star-formation rate (SFR) of the satellite galaxies. X-ray
observations of \MRC\ are inconclusive about the existent of an
extended X-ray atmosphere \citep[cf. ][the extended X-ray emission
could be due to inverse Compton or shocks generated by the passage
of the radio jets]{carilli02, pentericci00}.  \citet{pentericci00},
favoring the existence of a large thermal hot X-ray emitting atmosphere,
conclude that \MRC\ has many of the necessary ingredients of a forming
galaxy cluster, i.e. an irregular velocity distribution of the \Lya\
emitting galaxies, an over-density of galaxies, a massive central galaxy,
and a hot X-ray halo.  Due to the large number of companion galaxies
surrounding the \HzRG\ being analogous to ``flies caught in a spiders
web'', \MRC\ was dubbed the \Spider\ \citep{miley06}.

Multi-wavelength photometry, including the infrared continuum emission
\citep{stevens03, greve06, debreuck10}, imply an extremely high
SFR in the \Spider .  Fitting an active galactic nucleus (AGN) and
starburst component SED to the mid- to far-IR SED, \cite{seymour12}
find a SFR for the starburst component of
$1390\pm150$\,M$_{\odot}$\,yr$^{-1}$ \citep[see also ][]{ogle12}.
This should be compared with the rest-frame UV estimates of only a
couple 100 M$_{\odot}$\,yr$^{-1}$ (even with an extinction correction)
by \cite{hatch08}, who emphasise how deeply embedded the majority of
the intense star formation is in \MRC.  Although the star formation
is intense, the characteristics of the vigorous outflow ($\dot{\rm
M}\gtrsim400$\,M$_{\odot}$ yr$^{-1}$) observed in the optical emission
line gas suggest that this is predominately driven by the radio jet
\citep{nesvadba06}.

Beyond the companion galaxies, the diffuse stellar and gaseous
environment of \MRC\ on larger scales is also fascinatingly complex.
\MRC\ has a significant amount of diffuse UV intergalactic light (IGL)
within 60\,kpc of the radio galaxy indicating on-going star formation in
the circum-galactic environment \citep{hatch08}. This diffuse light is
embedded in a large ($\sim100$\,kpc in diameter) Ly$\alpha$ emitting halo
\citep{pentericci97}. Using semi-analytical models, \cite{hatch08} ruled
out the possibility that the observed circum-galactic light originates
from unresolved, low-mass satellite galaxies.  Spectra extracted at the
position of the central \HzRG, of a nearby galaxy and the IGL,
show \Lya\ emission lines with absorption troughs super imposed,
suggesting the presence of warm neutral gas mixed with the ionised gas
surrounding the \HzRG\ \citep{pentericci97, hatch08}.  This leads to the
conclusion that the large \Lya\ halo emission is powered not only by the
extended and diffuse star-formation \citep{pentericci98, miley06, hatch08}
but also by AGN photoionisation and shock heating \citep{nesvadba06}.

The influence of the radio jet from the AGN is seen in the VLA
observations, which reveal a bend in the western string of clumps detected
$\sim20$\,kpc from the core towards the south-west.  \Lya\ line emitting
gas has a bright spot associated with this bend and the two hotspots,
implying the presence of a massive cloud of gas deflecting the
radio jet and causing these features \citep{pentericci97,lonsdale86}.
Several different models exist for how gas might deflect radio jets, such
as the jet drilling into a gas cloud where it blows a bubble in the hot
plasma \citep{lonsdale86} or through the counter pressure generated by
oblique reverse shocks in the cloud generated by a jet-cloud interaction
\citep{bicknell98}.  Given this situation, \cite{pentericci97}
argue that the relation between the radio emitting knots in the jet
and the high surface brightness \Lya\ halo emission, especially where
the eastern jet bends, suggest an interaction between the jet and the
ambient gas in the halo of \MRC. However, the origin of the \Lya\
emitting gas reservoir is still uncertain.

This high a SFR of the \HzRG\ means that a significant molecular gas
reservoir fuelling the star formation must be present.  \cite{emonts13}
probe the diffuse extended molecular gas reservoir using the CO(1--0)
line.  They find that there is approximately $6\times10^{10}$
M$_{\odot}$ of cold \Htwo\ gas over a scale of 10s of kpc surrounding the
\HzRG\ (also Emonts et al. 2016, in prep.).  The kinematics
of the cold molecular gas is relatively quiescent (Emonts et
al. 2016, in prep.). More surprisingly, \cite{ogle12} detect the 0-0 S(3)
rotational line of \Htwo\ in \MRC. The strength of the line,
allowing for a range of plausible excitation temperatures of \Htwo, imply
warm ($T>300$\,K) \Htwo\ masses of order  $10^{7}$ to $10^{9}$
M$_{\odot}$.  While the large {\it Spitzer} beam does not allow to spatially resolve the \Htwo\ emission, a plausible interpretation of the relatively large
mass of warm \Htwo\ gas in \MRC\ is that a fraction of the jet
energy is being dissipated as supersonic turbulence and shocks in the
dense gas in the immediate environment of the AGN.

Though the optically thick CO(1--0) emission line is a good tracer of
the diffuse molecular gas, the emission lines from neutral carbon (\CIn)
are arguably even better.  The \CIn\ lines have critical densities similar
to the low-$J$ CO lines, meaning that they probe the same phases of the
molecular gas.  As they are both optically thin, they therefore probe
higher column densities than CO \citep{papadopoulos04}.  While the \CIn\
lines are good tracers of the diffuse molecular gas, they are poor tracers
of the very dense star forming gas.  Molecular lines from e.g. \water,
HCN and CS, have a much higher critical density and therefore probe
the dense molecular star forming phase.  \cite{omont13} find a relation
between the far-infrared (FIR) and \water\ luminosities for a sample of
high-$z$ starburst galaxies.  The \water\ detections for this sample are
all associated with underlying FIR emission, implying that the \water\
emission traces star forming regions.  However, the \water\
molecules can also be excited in the dissipation of supersonic turbulence
in molecular gas or by slow shocks \citep[e.g.][]{flower10}. In
the case of purely shock excited \water, it is unlikely that underlying
FIR emission would be detected in regions of strong \water\ emission
\citep[e.g.][]{goicoechea15}.

Motivated to determine the energy source and distribution of
the strong dissipation as possibly observed through \Htwo\ emission
and determining the state of the molecular gas in \MRC, we proposed
for ALMA observations in Cycle 1. In this paper, we present our
results for the observed \dustcont\ continuum emission
(rest-frame $\sim$740 GHz), \water\ $2_{11}-2_{02}$ transition
(at $\nu_{\rm rest} = 752.03$\,GHz, which is hereafter often refereed to as \water),
[C{\small I}] $^3P_2\,\to\,^3$$P_1$ fine structure emission line (at
$\nu_{\rm rest} = 809.34$\,GHz, which is hereafter referred to as \CI) and \CO\ emission
line observations towards the \Spider. We find strong \dustcont\
continuum emission at the position of the \HzRG.  Near both radio lobes,
we  detect emission from \water, the first spatially resolved detection
of \water\ in a high-$z$ un-lensed galaxy.  We also detect strong \CI\
emission blended with weak \CO\ emission at the position of the \HzRG.
In \S~\ref{sec:obs} we present our Atacama Large Milimeter/submilimeter
Array (ALMA) submm observations. The results of these observations
are given in \S~\ref{sec:res}. We analyse and discuss them in
\S~\ref{sec:ana} and summarise our conclusions in \S~\ref{sec:con}. We
assume H$_0=73$\,km/s/Mpc, $\Omega_M=0.27$, and $\Omega_{\Lambda}=0.73$,
which implies a scale of 8.172\,kpc/\arcsec\ at $z=2.161$.

\begin{figure}
\centering
\includegraphics[trim=1cm 9.6cm 14.2cm 9.8cm, clip=true,scale=1.04,angle=90]{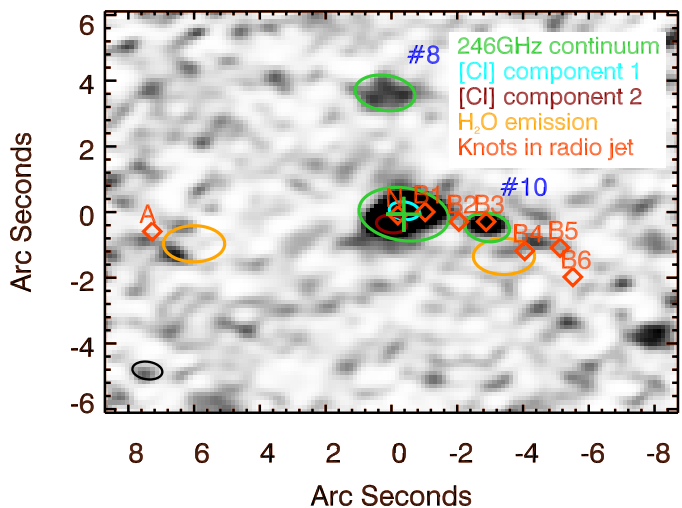}
\caption{{\small Overview of the spatial distribution of the detected
components. The natural weighted \dustcont\ continuum map is in grey
scale and the two \CI\ components 1 and 2 are marked with the blue and
red ellipses, respectively. The two \water\ detections are marked with
orange ellipses, and the \dustcont\ continuum components marked with
green ellipses. The sizes of the ellipses represent the extractions
used for the photometry.  The knots in the radio jet are marked with
red-orange diamonds and labeled according to \cite{pentericci97}.
The numbers correspond to the numbering in \citealt{kuiper11}. The ALMA
beam is shown as a black ellipse in the lower left corner.}}
\label{fig:overview}
\end{figure}

\section{Observations}\label{sec:obs}

\subsection{ALMA observations}

The ALMA cycle 1 Band 6 observations were carried out on 2014 April
27 for 49\,min on-source time with 36 working antennas.  The four
1.875\,GHz spectral windows were tuned to cover the frequency ranges
$237.3-240.9$\,GHz and $252.6-256.7$\,GHz.  We used the supplied Common
Astronomy Software Applications (CASA) calibration script to produce
the data cube, continuum map and moment-0 maps. The quasar J1146-2859
was used as a bandpass calibrator, and the UV range was well covered
within 400\,k$\lambda$.

We made a natural weighted map (Briggs robust parameter of 2),
which exhibits the highest S/N, at the expense of a lower
spatial resolution.  The frequency range between $254.9-265.7$\,GHz
(i.e. half of the upper side band) is dominated by strong \CI\ and \CO\
line emission and is therefore not included in the continuum map.  This
results in a continuum map with a synthesised beam of $0\farcs 69 \times
0 \farcs 44$ with PA~$89.3^{\circ}$ and an RMS of 50\,$\mu$Jy\text{/beam}.
We corrected the continuum map for the primary beam of 26\farcs3.

The \water\ emission is spatially offset from the continuum emission
and a continuum subtraction is therefore not required, as it would
only unnecessarily add noise. The \water\ observations are therefore
not continuum subtracted.  The \CI\ and \CO\ emission lines are within
the frequency range at $z=2.1606$, and we subtract the continuum in the
UV-plane by fitting a first order polynomial to the line free channels.
The \CI\ emission is so broad that it leaves no line free channels in
spectral window 0.  We therefore fit the continuum to spectral window
1 which shows no signs of line emission.  We bin the \CI\ line data
to 20\,km/s which has an RMS of 0.5\,mJy and the \water\ line data to
60\,km/s which has an RMS of 0.3\,mJy.  We primary beam correct the
\water\ line data, as we find \water\ emission $\sim6\farcs5$ from the
phase centre, where the sensitivity is at 87\%.

\section{Results}\label{sec:res}

The \Spider\ is detected in both \dustcont\ continuum and \CI, \CO\
and \water\ line emission.  Figure~\ref{fig:overview} shows an overview
of the spatial distribution of the different components. The strong
\CI\ line emission is seen in two components separated by 0\farcs5.
The \CI\ \compI\ is centred at the position of the \HzRG\ (marked
with a blue ellipse in Fig.~\ref{fig:overview}) and the \CI\ \compII\
is located 0\farcs5 to the south-east (marked with a red ellipse
in Fig.~\ref{fig:overview}).  The \dustcont\ continuum emission
peaks at the position of the \HzRG\ (marked with green ellipses in
Fig.~\ref{fig:overview}), but shows an extension in the direction of
the \CI\ \compII.  Emission from \water\ is also detected
south-west of the \HzRG\ (marked with the orange ellipse south of the
\dustcont\ continuum component in Fig.~\ref{fig:overview}), which is
co-spatial with the bend in the radio jet \citep{carilli97}.  \water\
emission is detected east of the \HzRG\ - west of knot A in the radio jet
(marked with an orange ellipse to the left in Fig.~\ref{fig:overview}).
We now discuss each of the components separately. Table~\ref{table:lines}
lists the derived line parameters.  Throughout, we will adopt the \CI\
emission at the position of the radio core as the systemic redshift
$z=2.1606$.

\subsection{Continuum emission}\label{res:dust}

The continuum map contains bright \dustcont\ continuum emission at the
position of the \HzRG, $\sim20$\,kpc to the west and tentative emission
to the north (see Fig.~\ref{fig:overview}).

\subsubsection{The \HzRG}

The \dustcont\ continuum emission at the position of the \HzRG\ shows
an east-west elongation, in the same orientation as the radio source
(see Fig.~\ref{fig:overview}). A similar orientation was seen in
the SCUBA and LABOCA maps \citep{stevens03,dannerbauer14}, but those
extensions were on a much larger scale than the separation seen in
Fig.~\ref{fig:overview}. Though the continuum only has one peak,
the elongation suggests a two component system, similar to what is
seen in the \CI\ line emission (see \S~\ref{sec:CI}) .  Using the
peak positions from \compI\ and 2 in the \CI\ moment-0 maps, we fit a
double 2D Gaussian profile to the continuum by fixing the centres of the
Gaussians at the peak positions of the \CI\ emission line.  This results
in a continuum ratio for the two components of $\sim$5. Integrating the
fitted double 2D Gaussian yields the flux density of $1.78\pm0.29$\,mJy
(see Table~\ref{table:cont}).  This emission is probably dominated
by thermal dust emission, though we warn that a linear extrapolation
of the $S(8.2\,\text{GHz)} = 1.88$\,mJy \citet{pentericci97} and the
$S(36.5\,\text{GHz}) = 1.02$\,mJy (Emonts et al, in prep.) for the
core predicts a synchrotron contribution of 0.47\,mJy at \dustcont. Any
star-formation parameters derived directly from the \dustcont\ should therefore
be considered as an upper limit.

\begin{table*}      
\centering          
\begin{tabular}{l c c c c}
\hline\hline       
Component                   & \multicolumn{2}{c}{Position} & $S_{246\,\text{GHz}}$ & $z$\\
                                      &             RA & dec                  &     mJy                          &       \\ 
\hline                    
\HzRG                           & 11:40:48.34 & -26:29:08.656 & $1.78\pm0.29$            & $2.1606\pm0.0041$\\ 
\hline
companion \#8              & 11:40:48.38 & -26:29:04.940 & $<0.25$                        & $2.1437$\\                 
companion \#10            & 11:40:48.15 & -26:29:09.490 & $0.19\pm0.01$             & $2.1446$\\                 
\hline                    
Water (West)               &  11:40:48.82 & -26:29:09.58   & $<0.2$ \\
Water (East)                &  11:40:48.12 & -26:29:09.97  & $<0.2$ \\
\hline                  
\end{tabular}
\caption{{\small The peak positions and \dustcont\ continuum flux of
the \HzRG\ and companion \#8 and \#10. The continuum fluxes are calculated
by integrating under the fitted double 2D Gaussian profile. Companion \#8 is a tentative 
detection and therefore an upper limit. We take the $3\sigma$
upper limit of the \dustcont\ continuum emission at the positions of
the \water\ emission to the west and east to be three times the RMS.}}
\label{table:cont}
\end{table*}

\subsubsection{The companion sources}\label{res:compantheions}

\begin{figure}
\centering
\includegraphics[trim=1cm 0.7cm 6.8cm 18.7cm, clip=true,scale=1,angle=90]{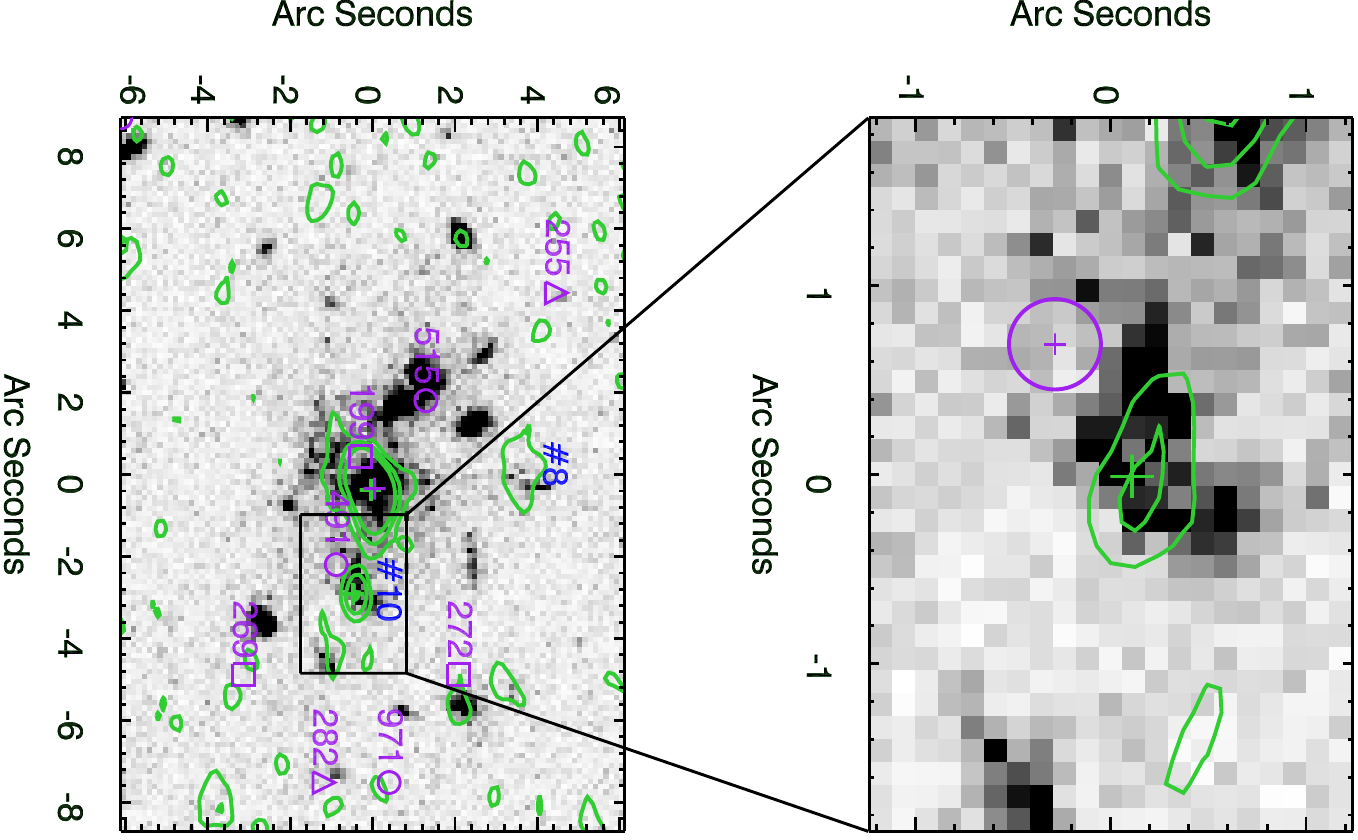}
\caption{{\small 
 \textit{Bottom panel}: \textit{HST} F814W image \citep{hatch08}
in grey scale overlaid with the natural weighted \dustcont\ continuum map
in green contours with levels of $1.5\sigma$, $3\sigma$ and $5\sigma$.
In the natural weighted continuum map we detect emission from companion
\#10 (and tentatively from \#8) of \citealt{kuiper11}.  The purple
cross marks the position of the nearby \Lya\ emitter at the position
of the \HzRG. The purple circles mark the positions of \Lya\ emitters,
the squares mark the position of \Ha\ and the triangles extremely red
objects within the \Lya\ halo of \MRC. The \Lya\ emitter \#491 is offset
by 0\farcs8 from the companion source seen in \dustcont\ continuum
emission west of the \HzRG. \textit{Top panel:} Zoom in of the region
around the companion sources. The purple circle marks the position of
the \Ha\ emitter \#491 \citep{kurk04a}, which is the close to the dust
continuum emission companion source.}}
\label{fig:HST}
\end{figure}

We detect \dustcont\ continuum emission from a bright companion
$\sim20$\,kpc to the west of the \HzRG\ (see Fig.~\ref{fig:overview}).
This companion is associated with knot B3 in the radio jet \citep[][see
Fig.~\ref{fig:overview}]{carilli97,pentericci97} and a small group of
galaxies (denoted D by \cite{pentericci97} and \#10 by \cite{kuiper11},
see Fig.~\ref{fig:HST}). This is also the position of a high surface
brightness region of Ly$\alpha$ emission \citep{pentericci97}. This
companion is also seen in e.g. \HST\ F814W imaging \citep{miley06} and at
other optical wavelength e.g. R-band \citep{pentericci97}. Companion \#10
is co-spatial with companion D in \cite{pentericci97} at RA = 11:40:48.14,
dec = -26.29.09.2, and is offset by only 0\farcs8 to the \Lya\ emitter
\#491 from \cite{kurk04a}.  This \Lya\ emitter at RA = 11:40:48.2, dec
= -26.29.09.5 (marked with the purple circle in Fig.~\ref{fig:HST} and
\ref{fig:HST}) is inside the \Lya\ halo of the \Spider\ \citep{kurk04a},
and so are three other \Lya\ emitters, three \Ha\ emitters and two
Extremely Red Objects (ERO) (also marked in Fig.~\ref{fig:HST}).
Fitting a 2D Gaussian profile to companion \#10 we find a flux density
of $0.19\pm0.01$\,mJy (see Table~\ref{table:cont}).

To examine the nature of the \dustcont\ emission of companion \#10,
we compared the flux density in the spectral windows not contaminated
by line emission.  The expected ratio between 238.28 and 253.48\,GHz
(i.e. the lower and upper side bands: LSB and USB) from thermal blackbody
radiation at 40\,K is 0.80, while the observed ratio is $1.19\pm0.22$. The
decreasing spectral slope with increasing frequency of \#10 is more
consistent with a synchrotron rather than a thermal dust origin.
Indeed, a straight extrapolation of the $S_{8.2\text{GHz}}=5.9$\,mJy
\citep{pentericci97} and the $S_{36.5\text{GHz}}=1.8$\,mJy (Emonts
et al., in prep) implies $S_{239\text{GHz}}=0.4$\,mJy.  Our observed
0.2\,mJy is therefore fully consistent with synchrotron emission, and
even allows for the expected spectral steepening at high frequencies.
Thus synchrotron dominated submm emission is consistent with companion
\#10 having very blue UV colours with no signs of a Balmer break or
significant extinction \citep{hatch09}.  On the other hand, it could also
be that the most obscured regions are not visible in the optical images,
but only the bluest, least obscured regions are.
Higher resolution dust continuum and a broad sub-millimetre wavelength
coverage can test these hypotheses.

\subsection{\CI\ and \CO\ line emission}
\subsubsection{The \HzRG}\label{sec:CI}
\begin{figure*}
\centering
\includegraphics[trim=0.65cm 0cm 5cm 0.9cm, clip=true,scale=0.69,angle=90]{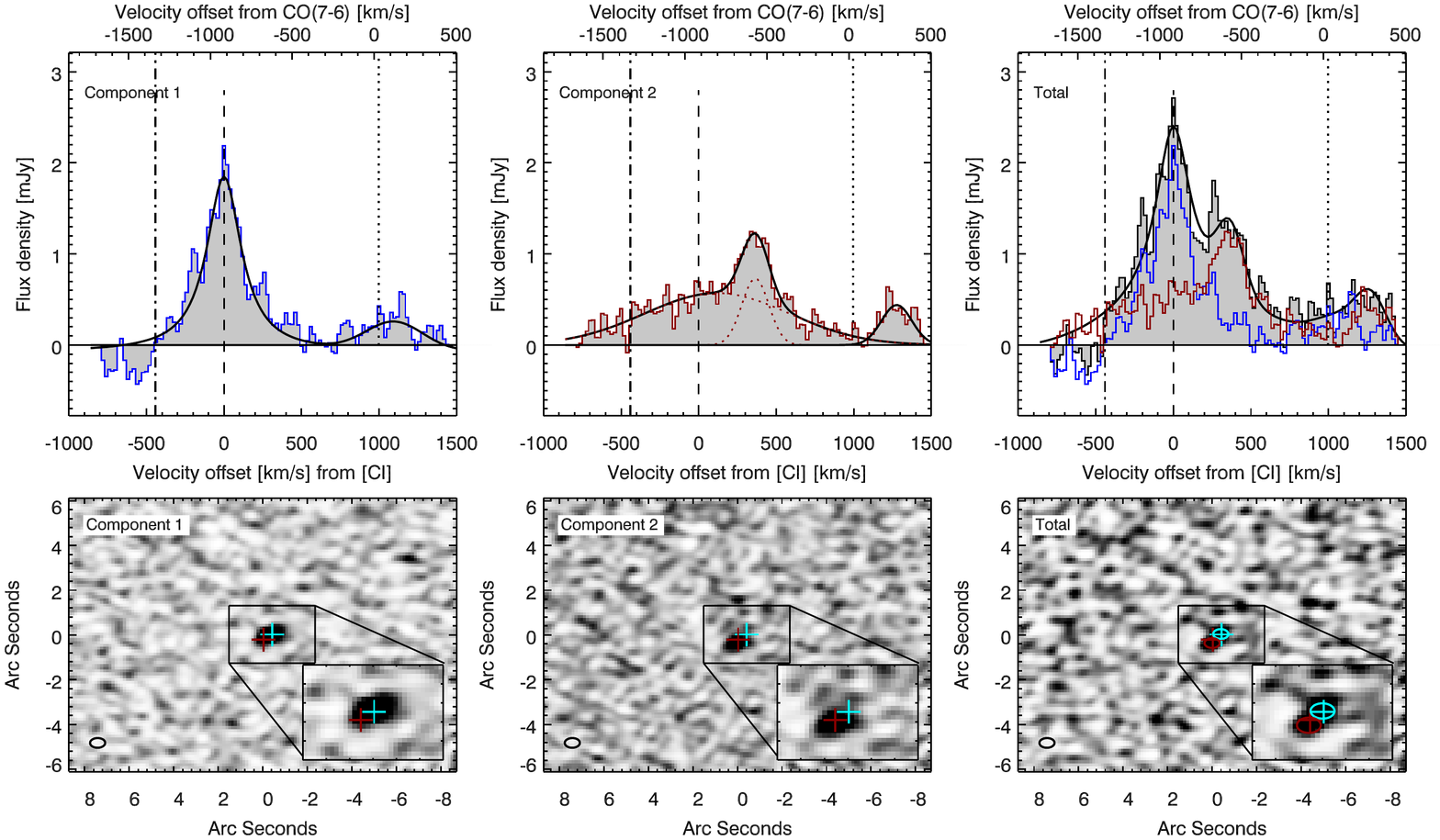}
\caption{{\small \CI\ spectra and moment-0 maps for \compI, \compII\
and the total \CI\ and \CO\ emission.  \textit{Top row:} The spectra
extracted from beam sized areas for \compI, \compII\ and the total. The
areas from which the spectra of \compI\ and \compII\ are extracted
do not overlap.  The Lorentzian profile for the \compI\ \CI\ line
(top left), the double Gaussian profile for the \compII\ \CI\ line
(top middle) and the single Gaussian fits for the two \CO\ lines are
over-plotted as black curves. The sum of the Lorentzian and Gaussian
profiles is over-plotted in black in the total spectrum (top right). The
dashed lines mark the 0-velocity of the \CI\ frequency at $z=2.1606$,
and the dotted lines mark the 0-velocity for the \CO\ frequency at the
same redshift. This redshift is in agreement with the redshift
determined from the CO(1--0) line \citep{emonts13}.  The dotted-dashed
line marks the 0-velocity of the \CI\ frequency at $z=2.156$ determined
from the \HeII~$\lambda$1640\AA\ line \citep[which as a non-resonant
line, should represent the systemic velocity of the AGN,][]{humphrey08}.
\textit{Bottom row:} The moment-0 maps of the \CI\ emission from \compI\
(bottom left), \compII\ (bottom middle) and the total (bottom right)
\CI\ emission and zoom-ins of the centres of the images. The total \CI\
moment-0 maps is overlaid with \CI\ line contours of \compI and 2. The
blue and red crosses indicate the peaks of the \CI\ emission of \compI\
(blue) and \compII\ (red).}}
\label{fig:CI}
\end{figure*}

Strong emission from the [C{\small I}] $^3P_2\,\to\,^3$$P_1$ fine
structure emission line is detected at the position of the \HzRG.
The \CI\ emission line has a double peaked velocity profile with peaks
at 0\,km/s and $\sim350$\,km/s (see top right in Fig.~\ref{fig:CI}).
Moment-0 maps of the channels containing the emission of the first
(see bottom left in Fig.~\ref{fig:CI}) and second (see bottom middle
in Fig.~\ref{fig:CI}) peaks reveal a spatial mis-alignment of the two
\CI\ peaks, which suggests two components. The emission corresponding
to the 0\,km/s peak is at the location of the \HzRG\ (\compI, marked
with a blue cross in Fig.~\ref{fig:CI}), while the $\sim350$\,km/s gas
is shifted 0\farcs5 to the south-east (\compII, marked with the red
cross in Fig.~\ref{fig:CI}).  A broad underlying components is visible in the spectrum
for \compII. Unfortunately, the spatial resolution of the data and the
broad underlying component do not allow for a clearer separation of the
two components.

The spectral velocity profile of \compI, with no overlap with
\compII\ (see Fig.~\ref{fig:CI} left), has a Lorentzian-shaped
profile, with a full-width at half-maxima (FWHM) of $270\pm15$\,km/s and a redshift of
$z=2.1606\pm0.0041$, which agrees with the redshift determined from the
CO(1--0) line \citep{emonts13}. We adopt this as the systemic redshift
as the \CI\ line has higher spectral resolution and S/N compared to the
\HeII\ $\lambda$1640\,\AA\ \citep{hatch08}.  At the expected frequency of
the \CO\ line, we detect a 3.5$\sigma$ Gaussian shaped \CO\ emission line
with FWHM of $435\pm85$\,km/s.  The broadness of the \CI\ line makes
the spectroscopic separation of the \CI\ and \CO\ lines difficult,
as very few line-free channels separate them.  To avoid the lines
contaminating each other, the velocity integrated line flux for the \CI\
line is calculated by integrating the fitted profile from $-$610\,km/s
to 530\,km/s, while the integrated \CO\ line flux is integrated from
690\,km/s to 1390\,km/s.  The \CO\ to \CI\ line luminosity ratio is 0.2.
The peak ratio of the two \CI\ components is $\sim3$ times lower than
the ratio of the \dustcont\ continuum at these positions, implying that
\compII\ is relatively brighter in \CI\ than in \dustcont\ emission.

The velocity profile of \compII\ shows evidence for two components,
one broad and one much narrower component (see Fig.~\ref{fig:CI}).
The best two-component Gaussian fit has a FWHM of $1100\pm65$\,km/s
for the broad, and $230\pm35$\,km/s for the narrow components.
The centre of the narrow component is shifted $\sim360$\,km/s redward
of the systemic velocity of the \CI\ line for \compI. The \CO\ emission
line is also detected for \compII, at a level of 4.8$\sigma$, which is
more significant than the \CO\ detection for \compI. The \CO\
line peaks at $\sim1275$\,km/s relative to the centre of the \CI\ line,
corresponding to a rest velocity for the \CO\ line of $\sim280$\,km/s.
The narrow component of the \CI\ line and the \CO\ line both have
an offset of 80\,km/s relative to the systemic redshift of component
1. Just as for \compI, separating the two lines spectroscopically is
difficult and we therefore calculate the integrated flux of the \CI\
line from $-$825\,km/s to 990\,km/s and the integrated flux of the \CO\
line from 1000\,km/s to 1490\,km/s.  We additionally derive the flux of
the narrow and broad components separately, by integrating the fitted
Gaussian profiles.  The \CO\ to \CI\ line luminosity ratio is 0.14,
lower than for \compI.  The two components have a \CI\ line peak ratio
of 1.7, and a \CO\ line peak ratio of 0.4.

The total spectrum of the two components (see Fig.~\ref{fig:CI} right)
clearly shows a double peaked \CI\ line and a broad \CO\ line.  We sum
the Lorentzian fit of \compI\ and the two Gaussian fits of \compII, which
results in a profile that fits well the observed integrated \CI\ line
of both components.  We likewise sum the two single Gaussians fitted to
the \CO\ lines in \compI\ and \compII. The full \CI\ plus \CO\ profile is
over-plotted on the full spectrum in the top right panel of Fig.~\ref{fig:CI}.
The \CO\ line in the total spectrum is even broader than for \compI\
and 2, making the separation even more difficult. We therefore calculate
the \CI\ integrated line flux from $-$590\,km/s to 700\,km/s and the \CO\
integrated line flux is integrated from 700\,km/s to 1390\,km/s. The \CO\
to \CI\ line luminosity ratio for the total spectrum is 0.2, the same
as for \compI.

\subsubsection{The companion source}

\begin{figure}
\centering
\includegraphics[trim=1cm 18.4cm 5.9cm 0.7cm, clip=true,scale=1,angle=90]{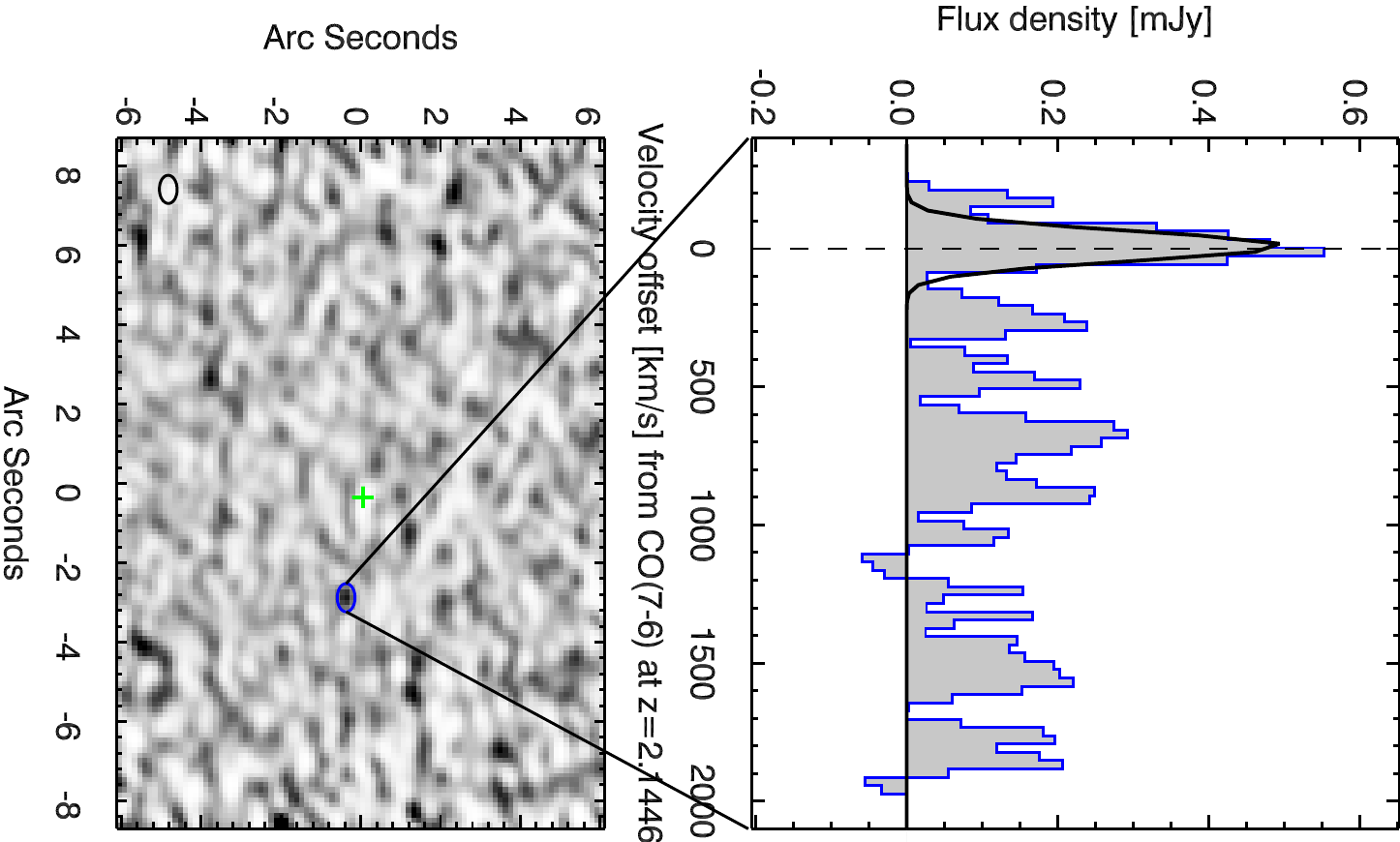}
\caption{{\small \textit{Bottom panel:} Moment-0 map of the \CO\ emission
at the position of the brightest companion.  \textit{Top panel:} The
spectrum extracted from the beam size area shown by the blue ellipse at
the position of the companion sources in the bottom panel and is binned
to 30\,km/s channels. The spectrum shows both the continuum and \CO\
line emission. The redshift of the line is consistent with the optical
$z=2.1446$ \citep{kuiper11} which was taken as zero velocity in the
spectrum.}}
\label{fig:COcomp10}
\end{figure}

We detect line emission in companion \#10 $\sim20$\,kpc to the west
of the \HzRG\ (see Fig.~\ref{fig:COcomp10}).  SINFONI spectroscopy of
this companion shows line emission from \Ha, \OII\ and \OIII, and that
companion \#10 has a velocity offset of $\sim-1360$\,km/s, corresponding
to a redshift of $z=2.1446$ \citep{kuiper11}.  Unfortunately, the \CI\
is shifted out of the observed band at this redshift and the \water\
line falls in a gap in the response of the band. However, the
\CO\ line is offset by 997\,km/s from the \CI\ line and therefore
still lies within the band at this redshift.  We identify the detected
emission line as \CO\ line emission from companion \#10
(see Fig.~\ref{fig:COcomp10}). It is therefore highly likely that
the \dustcont\ continuum emission that we detect at this position
is related to companion \#10.  This \CO\ line can be fitted with a
Gaussian profile with a FWHM of $130\pm20$\,km/s and the integrated
line flux from -170\,km/s to 150\,km/s of 0.08$\pm$0.01 Jy km/s
(Table~\ref{table:lines}).  We do not detect any line emission at
the position of the tentative \dustcont\ continuum companion \#8.

\begin{table*}  
\centering          
\begin{tabular}{l c c c c c c }
\hline\hline       
Transition & Frequency &  \multicolumn{2}{c}{Position} & $SdV$ & \multicolumn{2}{c}{FWHM} \\ 
                   &  GHz       &             RA & dec                   &  Jy\,km/s & km/s & km/s          \\ 
\hline
\multicolumn{7}{l}{\textbf{West}} \\
\water\ & 752.03 & 11:40:48.82 & -26:29:9.580 & $0.05\pm0.01$  & $230\pm50$ & --- \\ 
\multicolumn{7}{l}{\textbf{East}} \\
\water\ & 752.03 & 11:40:48.11 & -26:29:9.970 & $0.1\pm0.02$  & $350\pm70$ & --- \\
\multicolumn{7}{l}{\textbf{\compI}} \\
\CO\     & 806.65 & 11:40:48.33 & -26:29:8.582 & $0.13\pm0.02$  & $435\pm85$ & ---  \\ 
\CI\      & 809.34 &  11:40:48.33 & -26:29:8.582  & $0.66\pm0.02$   & $270\pm15$ & --- \\ 
\multicolumn{7}{l}{\textbf{\compII}} \\
\CO\     & 806.65 & 11:40:48.36 & -26:29:8.822 &  $0.11\pm0.01$  & $230\pm35$ & ---  \\
\CI\      & 809.34  & 11:40:48.36 & -26:29:8.822 &  $0.79\pm0.03^a$  & $1100\pm65$ & $205\pm20$ \\
\multicolumn{7}{l}{\textbf{\compI + \compII}} \\
\water\  & 752.03 & 11:40:48.33 & -26:29:8.582 & $<0.9^b$                &      ----            & ---\\
\CO\     & 806.65 & 11:40:48.33 & -26:29:8.582 &  $0.28\pm0.02$  & $720\pm110$ & --- \\
\CI\      & 809.34  & 11:40:48.33 & -26:29:8.582  &  $1.30\pm0.03$  & $420\pm30$ & $280\pm40$  \\
\hline
\multicolumn{7}{l}{\textbf{Companion \#10}} \\
\CO\     & 806.65 & 11:40:48.15 & -26:29:09.075 &  $0.08\pm0.01$  & $130\pm20$ & --- \\ 
\hline
\end{tabular}
\caption{{\small The \water, \CI\ and \CO\ emission line positions,
velocity integrated fluxes and FWHMs. Fluxes and fitted FWHMs are
given for \CI\ \compI, 2, total and the \water\ components. The
spectra are extracted within a synthesised beam size which for
the \CI\ observations is $0\farcs 72 \times 0 \farcs 45$ with
pa~$87.4^{\circ}$} and for the \water\ observations is $0\farcs 94
\times 0 \farcs 56$ with pa~$-84.1^{\circ}$.  $^a$The \CI\ flux is
composed of 0.63$\pm$0.03\,Jy\,km/s for the broad velocity gas, and
0.16$\pm$0.03\,Jy\,km/s for the narrow velocity gas.  $^b$The $3\sigma$
upper limit of the \water\ emission taking to be $3\times$ the RMS in
60\,km/s wide channels and assuming a width of the line to be that of
the \CO\ line.}
\label{table:lines}
\end{table*}

\subsection{\water\ line emission}

\begin{figure*}
\centering
\includegraphics[trim=0.65cm 0cm 5cm 0.7cm, clip=true,scale=0.69,angle=90]{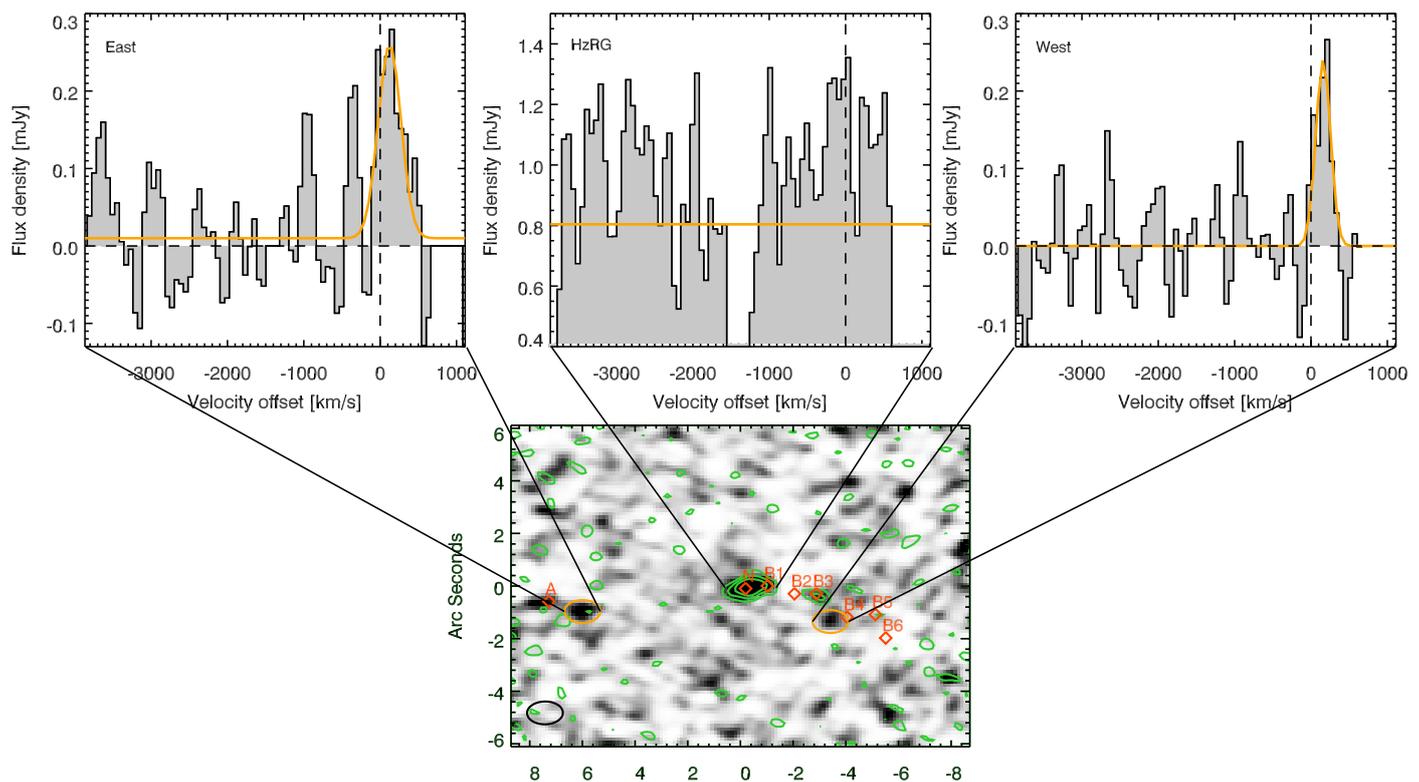}
\caption{{\small The non-continuum subtracted \water\ spectra for the
detection of the west and east and the non detection at the position
at the \HzRG\ and \water\ moment-0 map.  \textit{Top left panel:}
The \water\ line detected $\sim50$\,kpc to the east of the radio core,
has a $3.7\sigma$ significance. The emission is located due west of the
knot A in the radio jet. The fitted Gaussian is over-plotted in orange.
\textit{Top middle panel:} Spectrum at the position of the radio core,
shows no detection of \water\ emission -- only continuum emission. A
small separation between the two spectral windows results in the gap
in the continuum between $-1300$ and $-1500$\,km/s.  \textit{Top right
panel:} The \water\ line detected $\sim25$\,kpc to the west of the radio
core, showing a $4\sigma$ \water\ detection at the expected frequency
for $z=2.161$. The emission is located at the bend of the radio jet, B4
\citep{pentericci97}. The best fit Gaussian is over-plotted in orange.
\textit{Bottom panel:} Moment-0 map of the \water\ emission ({\it without}
continuum subtraction), overlaid with the \dustcont\ continuum emission
in green contours. The orange ellipses mark the \water\ emission and the
orange-red diamonds the positions of the knots in the radio jet given
by \citet{pentericci97}.}}
\label{fig:H2O}
\end{figure*}

We detect emission from the \water\ $2_{11}-2_{02}$ transition
$\sim25$\,kpc to the west and $\sim50$\,kpc to the east of the radio core
at the expected observed frequency for \water\ at $z=2.1606$.  The western
$4\sigma$ detection is located just south of the strongest \dustcont\
continuum companion, at the bend of the radio jet i.e. at radio knot B4
(see Fig.~\ref{fig:H2O}).  The  eastern $3.7\sigma$ detection is located
west of the radio knot A (see Fig.~\ref{fig:H2O}). To establish that this
emission is real we perform a 'Jackknife' test, cutting the observed
time in half. The two \water\ lines show up in both halves of the data
with a $\gtrsim2\sigma$ significance, which suggests that the
emission lines are real and not simply noise peaks. Fitting Gaussian
profiles to the emission lines result in FWHMs and relative velocities
of $230\pm50$  and 160$\pm$20\,km/s for the western detection
and $350\pm70$ and 125$\pm$30\,km/s for the eastern detection
respectively.  No \dustcont\ continuum emission is detected at the
positions of the two \water\ detections down to an RMS of 40\,$\mu$Jy,
and no \water\ emission line is detected at the position of the \HzRG\
(see Fig.~\ref{fig:H2O}) down to an RMS of 40\,$\mu$Jy in 60\,km/s wide
channels; assuming the 720\,km/s width of the \CO\ line from the total
spectrum, yields a $3\sigma$ upper limit of 0.9\,Jy\,km/s.

\section{Analysis and discussion}\label{sec:ana}

The lines we have detected in the \Spider\ are useful for a wide
range of gas diagnostics. The atomic forbidden line of carbon, \CI,
is a good tracer of relatively diffuse, low extinction molecular
gas \citep[e.g.][]{papadopoulos04}, as its line strength
is linearly proportional to the column density of molecular gas
\citep{glover15}. \CO\ emission is strong in dense, highly excited
optically thick molecular gas. The thermal dust continuum over
the range of a few 100 GHz represents the cooling of dust heated by the
intense stellar and AGN radiation fields within the \Spider.  The \water\
$2_{11}-2_{02}$ line is excited both in slow shocks ($10-40$\,km/s) in dense molecular gas \citep[10$^{3-5}$ cm$^{-3}$; ][]{flower10} or by
IR pumping \citep[e.g.][]{vanderwerf11}. Thus our data, in principle,
probe a wide range of conditions and heating/excitation mechanisms in
(relatively dense) molecular and atomic gas.

\subsection{Diffuse and dense molecular gas}

\subsubsection{Line velocity profiles}

\begin{figure*}
\centering
\includegraphics[trim=7.6cm 0.7cm 5.8cm 1.1cm, clip=true,scale=0.69,angle=90]{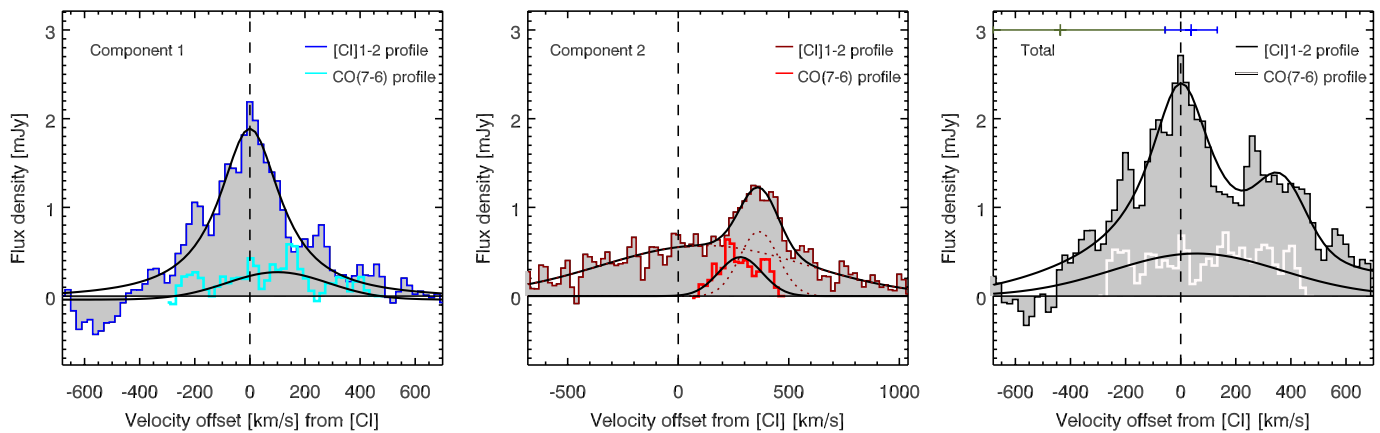}
\caption{{\small The \CI\ spectra (dark blue, dark red and black
histograms) for \compI, \compII\ and the total over-plotted with the
respective \CO\ lines (light blue, orange and white histograms) to compare
the velocity profiles. The fitted profiles from \S~\ref{sec:res} are
over-plotted as black curves, with the individual velocity components
as dotted lines. The narrow \CI\ component of \compII\ is offset by
365\,km/s from the systemic redshift. The bar above the velocity profiles
in the right plot, marked the \HeII\ redshift and errors (olive bar)
from \cite{roettgering97} and the CO(1--0) redshift and error (blue bar)
from \cite{emonts13}.}}
\label{fig:profiles}
\end{figure*}

Fitting of Lorentzian and Gaussian profiles for the emission line shows
that the \CI\ and \CO\ lines \compI\ and 2 have very different velocity
profiles.  Figure~\ref{fig:profiles} compares the velocity profiles of
the \CI\ and \CO\ emission lines for \compI\ (left), \compII\ (middle)
and the total (right). Both the \CI\ and \CO\ emission lines of \compI\
are broad, but though \CI\ and \CO\ do not trace exactly the same phases
of the molecular gas, the broadness of the two lines does suggest that the
two phases are related.  The velocity profiles of the \CI\ and \CO\ lines
for \compII\ show similarities between the narrow \CI\ component
and the \CO\ line. The broad \CI\ component is not seen in the
\CO\ line because of both limited spectral coverage of our observations
on the high velocity end of our bandpass, and limited S/N of this faint
line. The fact that the narrow velocity gas in component 2 is detected
in both \CI\ and \CO\ emission and with similar FWHM indicate that this
is a well-defined object. High spatial resolution observations would
be required to determine if this is a separate self-gravitating object,
or a kinematic feature within the same physical object.  The broadness
of the \CO\ emission in the total spectrum indicates that this emission
is dominated by the \CO\ emission from \compI.

The \CO\ lines for both \compI\ and 2, show an offset from the systemic
redshift of $\sim90-100$\,km/s.  A similar offset between the \CIn\
and CO lines is seen in a study of H1413+117 (the Cloverleaf Galaxy) by
\cite{weiss03}.  They conclude that the reason for the shift is unclear,
but that gravitational amplification should not alter the frequency of
the lines, unless the distribution of the \CIn\ and CO emitting gas
is different.  The \CIn\ and CO lines for the Cloverleaf Galaxy have
low S/N, \cite{weiss03} therefore conclude that high S/N observations
are necessary to confirm the offset and its origin, but that, if it is
genuine, it is most likely due to opacity effects.  Since we here see a
similar offset for the \Spider, which is an un-lensed source, differential
lensing cannot be the cause of this offset, however, differences in the
opacity of the two lines can.  The \CO\ line has a higher optical depth
than the \CI\ line ($\tau_{\text{[CI]}}=0.1$, \citealt{weiss03}), meaning
that the \CO\ emitted photons go through internal absorption and emission.
The line photons whose frequencies have been shifted away from the
systemic frequency of the \CO\ emission line, have a higher probability
of escaping, thus shifting the overall observed frequency.

The \Ha\ line for the \Spider\ was detected with SINFONI
\citep{nesvadba06} and  ISAAC \citep{humphrey08}.  The \Ha\
($\lambda$\,6550), \NII\ ($\lambda$\,6585) lines and \SII\
($\lambda\lambda$\,6718, 6733) doublet are spectrally very close and the
lines are therefore blended together. Both studies fit the blended
lines with multiple Gaussians. \cite{nesvadba06} finds FWHM = 14900\,km/s
for the AGN component and FWHM $<2400\,$km/s for the emission originating
from a region surrounding the AGN. This is the same order of magnitude
as what \cite{humphrey08} find by fitting a six Gaussian profiles,
including two \Ha\ components.  These two \Ha\ components consist of a
transmitted broad line region from the AGN (FWHM $=13900\pm500$\,km/s),
and a component with FWHM = 1200$\pm$80\,km/s. The relatively low
spectral resolution of the ISAAC spectrum does not allow for a further
de-blending of the lines.  However, the \Ha\ line profile does show the
presence of a broad and narrow component, similar to what we observe in
the \CI\ emission of \compII.  This strengthens the idea that the \CI\
emission traces the bulk of the gas in the galaxy, which is consistent
with the \CIn\ emitting gas being distributed throughout and generally
tracing gas of moderate extinctions and relatively low density molecular
gas \citep{papadopoulos04}.

\subsubsection{Line ratios}
\begin{table*}      
\centering
\begin{tabular}{l c c c c l}
\hline\hline       
Source                      & $z$      & $\mu$ & $SdV_{\text{\CO}}$ & $SdV_{\text{[CI]}}$ & reference \\
                                 &             &            & [Jy km/s]                 &       [Jy km/s]           &                 \\ 
\hline
SDP.11                      & 1.7860 & 18       & $18\pm14$     & $31\pm14$      &\cite{lupu12,ferkinhoff14}\\ 
SDP17.b                   & 2.3080 & 4.3      &$11\pm7$        & $13\pm7$        &\cite{lupu12}\\ 
SMMJ2135-0102      & 2.3259 & 32.5    & $12.6\pm0.6$ & $16.2\pm0.6$  &\cite{danielson11}\\
SMMJ16359+6612   & 2.5160 & 22       & $3.3\pm1.4$   & $1.6\pm0.3$    &\cite{kneib05, walter11}\\
H1413+117               & 2.5585 & 11       & $44.6\pm3.1$ & $5.2\pm0.3$    &\cite{alloin97, weiss03}\\
SMMJ14011+0252   & 2.5650 & 23       & $3.2\pm0.5$   & $3.1\pm0.3$    &\cite{downes03, walter11}\\
SDP.81                      & 3.0370 & 9.5     & $12\pm4$        & $<6.5$            &\cite{lupu12}\\ 
MM18423+5938       & 3.9296 & 12       & $3.9\pm0.5$   & $4.2\pm0.8$    & \cite{lestrade10}\\ 
ID141                        & 4.2430 & 10-30 & $6.5\pm1.4$    & $3.4\pm1.1$   &\cite{cox11}\\ 
HFLS3                      & 6.3369 &  2.2     & $2.2\pm0.3$   & $0.5\pm0.4$    & \cite{riechers13}\\ 
\hline\hline
                                 &             &            & $SdV_{\text{\water}}$ & $\mu L_{\text{IR}}$ & reference \\
                                 &             &            &          [Jy km/s]             &   [$10^{13}\,L_{\odot}$]   &                 \\
\hline
SDP9                        & 1.574   &   8.5    & $14.4\pm1.1$              & $4.4$    &   \cite{lupu12} \\
NAv1144                   &  2.202  &   5.3    & $7.5\pm 0.9$               &  $5.7$    &   \cite{omont13}\\
G15v2779                 & 4.244   &   4.1    & $4.1\pm0.6$                & $8.5$    &   \cite{omont13}\\
APM08279+5255      & 3.912   &   4.0    & $<1.1$                         & $20$     &    \cite{vanderwerf11, walter11}\\
HFLS3                      & 6.3369 &  2.2     & $2.6\pm0.8$                & $4.2$    &   \cite{riechers13} \\
\hline                  
\end{tabular}
\caption{{\small SMGs and QSOs from the literature with published \CI,
\CO\ and \water\ detections along with published IR luminosities used
for comparison to the \Spider. }}
\label{table:compare}
\end{table*}

\begin{figure}
\centering
\includegraphics[trim=7.7cm 9.6cm 5.9cm 10cm, clip=true,scale=1.05,angle=90]{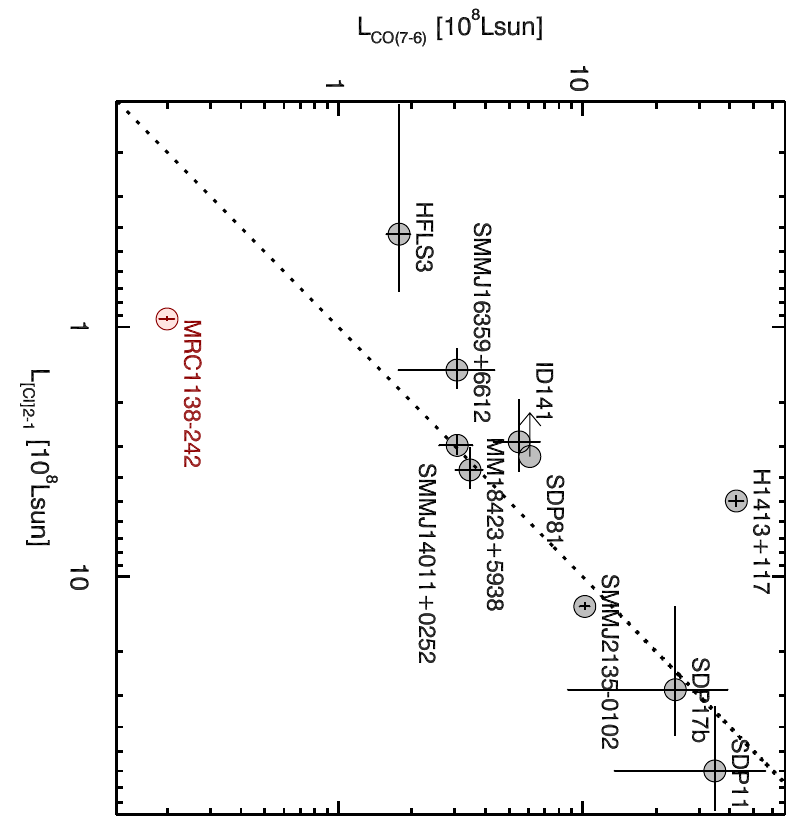}
\caption{{\small The \CO\ vs the \CI\ luminosity for a sample of high-$z$
SMGs and QSO (grey circles) and the \Spider\ (red circles).  The dotted
curve is the 1-1 relation. }}
\label{fig:ratios}
\end{figure}

The utility of using the \CIn\ emission lines as tracers of the
\Htwo\ column density has been examined by \cite{papadopoulos04}
by considering the \CI\ to CO line ratio.  The \CIn\ lines,
\CIn$^3P_1\to^3P_0$ (\CIn1--0) and \CIn\ $^3P_2\to^3P_1$ (\CI), have
critical densities of $n_{\text{cr,[CI]1-0}}\sim500$\,cm$^{-3}$ and
$n_{\text{cr,[CI]2-1}}\sim10^3$\,cm$^{-3}$, similar to that of CO(1--0)
and CO(2--1), which are often used as \Htwo\ tracers.  The \CIn\
lines have low to moderate optical depths ($\tau\approx0.1-1$),
which means that these lines have the advantage, compared to CO, of
tracing higher column-density cold diffuse molecular gas.  The \CIn\
lines are also more easily excited.  These properties make the \CIn\
lines more direct tracers of the molecular gas mass than the optically
thick $^{12}$CO lines.  Though the \Htwo-tracing capability of \CIn\
decreases for low metallicities,  \cite{papadopoulos04} conclude it is
still a better \Htwo\ tracer than $^{12}$CO \citep[see also ][]{glover15}.

The ground-state transition, \CIn1-0, is the most direct \Htwo\
gas mass tracer of the two \CIn\ lines because it is (relatively) less
sensitive to the excitation conditions of the gas. However, the \CI\
is still a much better tracer of the \Htwo\ gas than the \CO\ line at
similar frequency.  The much higher critical density of the \CO\ line of
$n_{\text{cr,CO(7-6)}}\sim3\times10^6$\,cm$^{-3}$ and higher excitation
energy of $E_{76}/k\sim155$\,K, compared to $E_{21}/k\sim62$\,K for
the \CI\ line, means that the \CI\ and \CO\ lines are unlikely to trace
exactly the same molecular gas phases.  The  \CO\ line is likely tracing
the higher density, more highly excited molecular gas.

We searched the literature for high-$z$ sources with observations
of both the \CI\ and \CO\ emission lines and determined IR luminosities,
and find nine lensed sub-millimetre galaxies (SMG) and QSOs
(see Table~\ref{table:compare}). To compare the lensed sources
and the un-lensed \Spider, which have different luminosities, we
normalise the \CI\ and \CO\ line luminosities by the IR luminosity,
which cancels out the lensing magnification for the lensed sources
(assuming there is no differential lensing between the IR and line
emission). Figure~\ref{fig:ratios} shows a positive correlation between
these normalised line luminosities.  The only outlier in the comparison
sample is the highly lensed Cloverleaf galaxy \citep[][]{alloin97,
weiss03, vanderwerf11}. The \Spider\ is an outlier, but on the low side
of the correlation.

\subsubsection{Possible causes of observed line ratio}

The \Spider\ is, like the Cloverleaf galaxy, an outlier compared to
lensed galaxies.  This unusually low \CO/\CI\ line ratio can be due
to the \Spider\ being an extraordinary \HzRG, with a highly unusual
ratio between the diffuse and dense molecular gas, or due to impact of
cosmic-ray heating and ionisation, or differential lensing.

Comparing $L^{\prime}_{\rm CO(7-6)}=1.2\times 10^{10}$\,K\,km/s\,pc$^2$
with $L^{\prime}_{\rm CO(1-0)}=6.5\times 10^{10}$\,K\,km/s\,pc$^2$
\citep{emonts13}, we find that the \CO\ emission is only 18\%
of the value for thermalised gas. This indicates that the molecular
gas phase is not thermalised and dominated by the cold and diffuse gas,
traced by the \CI\ emission line \footnote{Note, however, that
the high spatial resolution of the ALMA observation means that up to 2/3
of the \CO\ emission may not be detected due to lack of sensitivity to
very extended, low surface-brightness line emission (Emonts et al. 2016,
in prep).}.  Alternatively, a fraction of the CO in the \Spider\ may be
dissociated by cosmic-rays, increasing the fraction of atomic carbon in
the molecular gas without strongly affecting the excitation of \CI\ or
the ionisation state of carbon \citep{bisbas15}.  In the circum-nuclear
region of a powerful radio galaxy, it is likely that cosmic-rays
may have a significant impact on the nature of the molecular gas.
More observations of other atomic and molecular species in the \Spider\
are necessary to investigate this quantitatively.

However, since the comparison sample are all lensed galaxies, the
difference could also be caused by differential lensing in the comparison
sample.  Dense (traced by \CO) and diffuse (traced by \CI) gas in galaxies
is may be subject to differential lensing, where the emission from
the densest gas have a higher overall magnification than the diffuse gas.
\cite{serjeant12} showed that for lensing magnifications $\mu>2$, the CO
ladder can be strongly distorted by differential lensing.  Since all the
SMGs and QSOs listed in Table~\ref{table:compare} have a magnification
$\mu>2$ it may be that their line ratios are all influenced by the affect
of differential lensing.  If so, this implies that the \CO\ emission
likely has a higher magnification than the \CI\ emission, which would mean
that the intrinsic \CO/\CI\ ratio would be lower. We note however,
that although some of the data are limited by S/N and/or
low resolution, the line profiles of \CO\ and \CI\ are consistent with each other (see
references listed in Table~\ref{table:compare}).  While not conclusive,
the similarity of their dynamics suggests that the gas probed by \CO\
and \CI\ follow the same over all dynamics and are perhaps not strongly
affected by differential lensing.

We are planning to extend our ALMA \CIn\ observations to eight \HzRG s
to verify if this low CO/\CIn\ ratio is a common feature and also more
detailed studies of the \Spider\ to investigate this puzzle.

\subsection{Cooling of the post-shock gas due to slow shocks in molecular halos}

\begin{figure}
\centering
\includegraphics[trim=7.8cm 0.6cm 5.9cm 18.7cm, clip=true,scale=1.05,angle=90]{Figures/ratios.pdf}
\caption{{\small The $L_{\text{\water}}/L_{\text{IR}}$-ratio versus
{\it intrinsic} $L_{\text{IR}}$ for the \water\ detections from
\citealt{omont13} along with the \water\ detection for the \Spider. The
grey circles are sources from \citealt{omont13} with directed detections
of the $2_{02}-1_{11}$ line and the grey triangles are sources with
$2_{11}-2_{02}$ detection scaled using Mrk231 as a template to get an
estimate on the $2_{02}-1_{11}$ luminosity. The \Spider\ is marked with
the red point. We here plot the summed \water\ emission of the Western
and Eastern detection, as the IR luminosity is integrated over the
full source.}}
\label{fig:ratiosh2o}
\end{figure}

Interestingly, the two regions of \water\ emission that we detect do
not appear to be associated with significant dust continuum emission.
This is puzzling as a close association between a dense gas
tracer such as \water\ and dust is expected, as the excitation of the
\water\ emission in other high-$z$ (lensed) sources has been attributed
to IR pumping \citep{vanderwerf11, omont13}.  Moreover, no \water\
emission is detected at the position of the \HzRG\ (Fig.~\ref{fig:H2O}),
which is surprising given its strong IR emission and high rate of
mechanical energy injection via the radio jets (Fig.~\ref{fig:H2O}).
Specifically, \cite{omont13} detected \water\ for seven
high-$z$ lensed sources from the H-ATLAS survey. Five were detected in
\water~$2_{11}-2_{02}$ emission, and three in \water~$2_{02}-1_{11}$
emission.  Comparing these \water\ detections to the 8-1000\,$\mu$m IR
luminosity \cite{omont13} find a correlation between the IR and \water\
line luminosity of $L_{\text{\water}}=L_{\text{IR}}^{1.1-1.3}$, with the
exact exponent depending on which of the two \water\ lines is used in
the analysis.  From this relation, they conclude that IR pumping is the
most likely mechanism for exciting the water molecules and their line
emission. This is consistent with \water\ tracing dense gas surrounding
regions of intense star formation. Furthermore, a similar conclusion was
reached in a more detailed study of the \water\ emission from the lensed
QSO, APM\,08279+5255 at $z=3.9$ \citep{vanderwerf11}, and by
\cite{yang13}, who detected \water\ emission lines in 45 low-$z$ galaxies.

Figure~\ref{fig:ratiosh2o} plots $L_{\text{\water}}/L_{\text{IR}}$
versus $L_{\text{IR}}$ and shows an anti-correlation for IR luminous
high-$z$ lensed sources \citep{omont13, vanderwerf11}.  SDP.81 is an
outlier in this relation, most likely due to an underestimating of the
\water\ luminosity because of low S/N in the spectrum, and the large
extent of the source \citep{omont13}. We plot the \Spider\ as the
sum of the \water\ luminosities of the western and eastern detections
and the total infrared luminosity.  The \water\ emission originates
from two small regions, and any more widespread emission would not
be detected due to the lack of sensitivity of our ALMA data
on larger spatial scales. The IRAM Plateau de Bure observations of the
comparison sample of \citet{omont13} have 2 to 7 times larger synthesised
beam sizes, and may be more sensitive to emission at larger scales.
The $L_{\text{\water}}/L_{\text{FIR}}$ ratio for the \Spider\ is therefore
a lower limit and the \Spider\ could therefore be an outlier in
this trend. The lack of \dustcont\ continuum emission at the \water\
positions, which would imply a much larger deviation from the trend,
implies that the emission is not pumped by the IR emission.  The \water\
detection is spatially located at the bend in the radio continuum emission
(knot B3) and upstream to the radio jet.  Similarly, the eastern \water\
detection upstream from knot A in the radio emission, is likely associated
with the jet which is projected onto the plane of the sky.  

The most straight-forward explanation for the \water\ emission at
these positions, the lack of significant associated \dustcont\
continuum emission, and the offset from the anti-correlation
in Fig.~\ref{fig:ratiosh2o}, is that the \water\ emitting gas is
shock-heated and represents strong cooling in post-shock heated gas by
the passage of the radio jet. \water\ is one of the main carriers of
oxygen along with CO, but has many more radiative degrees of freedom,
meaning that it can lose energy efficiently under a wide range of (low
energy) excitation conditions and thus contribute significantly to the
overall cooling of dense molecular gas, in particular after the shock
impact \citep{flower10}. Given the small offset of the line peaks from
the systemic [CI]2–1 redshift, only 100 to 200\,km/s, and the
average line width of  $290$\,km/s of the two \water\ lines,
it is likely that the width of the lines represents the dispersion of
the clouds, and that mechanical energy is being dissipated within the
dense molecular gas. Not only \water , but also the detection of strong
mid-infrared \Htwo\ emission \citep{ogle12}, suggest excitation of
molecules in low-velocity shocks \citep[see discussion in ][]{guillard09,
appleton13}. Moreover, we emphasise that due to the position of the
\water\ emission in our bandpass, higher negative offset velocity
emission, such as that from companion \# 10 would not lie within our
bandpass \citep{kuiper11}.

Furthermore, it is unlikely that the dense molecular gas has been
entrained during the passage of the radio jet and lifted out of the
radio galaxy or any of its companions.  There are several reasons why
this is unlikely.  First, both regions of \water\ emission are almost at
the systemic velocity of the radio galaxy (as determined from CO(1-0),
\CI\, or \CO).  The \water\ lines have systemic velocities that are
significantly lower than any of the companion galaxies with determined
redshifts near the sites of \water\ emission.  Second, the cooling
time of the dense shock gas is of-order $10^3-10^4$\,yr for a
20\,km/s shock driven into a gas with a density of $10^5$\,cm$^{-3}$
\citep{flower10}, much shorter than any plausible dynamical time. These
facts suggest that the dense molecular gas is forming in situ in
the post-shocked gas after the passage of the radio jet.  It is not
surprising, within this context, to see the emission near the bend in the
jet and near the terminal hot spot of the western radio lobe as these
are sites where the interaction between the jet and ambient medium are
likely to be strong.  In such strong interaction zones, dissipation of
energy in a turbulent cascade is important \citep[e.g.][]{guillard10}. The
eastern terminal lobe of the jet is, for example, the working surface
of the jet as it propagates outwards, just the region where you would
expect strong dissipation of the bulk kinetic energy of the jet.

The molecular gas likely formed \textit{in situ} in the post-shock gas,
which has important implications for the state of the circum-galactic
halo and cluster gas. 
Recent observations of the \Spider\ using ATCA discovered a large region of
CO(1--0) surrounding the \HzRG\ with strong emission near some of the radio knots and 
lobes suggesting that the passage of the radio jet enhances the molecular column densities (Emonts et al. 2016, in prep.). 
Finding strong \water\ line emission
near the radio jet provides evidence of the formation of molecular gas
in the post-shocked gas of the expanding radio source \citep[see ][
for a discussion of this mechanism in the violent collision in the
Stephan's Quintet group]{guillard10, guillard12, appleton13}.  This is
also substantiated by the low relative velocities of the \water\ emission which
are consistent with the low velocities observed in the extended diffuse
CO emission.  Such a mechanism might also explain the distribution of the
CO emission observed in local clusters \citep{salome11} and around other
\HzRG s \citep[][ and references therein]{emonts14}.  Finding \water\
emission indicating strong dissipation of the bulk kinetic energy of
expanding radio jets provides a direct link between the jets and the
formation of molecular gas and stars in the halo of the radio galaxy.

\subsection{Mass Estimates}

As argued above, the \CI\ line is a better tracer of the cold molecular
gas than the high-$J$ CO lines, due to their lower critical density. We
therefore use the \CI\ to estimate the total molecular gas mass, following
the approach of \cite{papadopoulos04} and \cite{alaghband-zadeh13}:

\begin{align}\label{eq:Mh2}
M_{\text{\Htwo}} = 1375.8\,D_L^2\,(1+z)^{-1}\,{\left[\frac{X_{\text{[CI]}}}{10^{-5}}\right]}^{-1} \\
\nonumber
\times{\left[\frac{A_{21}}{10^{-7}\text{s}^{-1}}\right]}^{-1}Q_{21}^{-1} \left[\frac{S_{\text{[CI]2-1}}dV}{\text{Jy\,km\,s}^{-1}}\right],
\end{align}

\noindent
where $X_{\text{[CI]}}$ is the \CIn-to-\Htwo\ abundance for which
we adopt the literature-standard of $3\times10^{-5}$, $Q_{21}$
is the density dependent excitation factor for which we assume the
median value of 0.57 \citep[see][]{papadopoulos04}, and $A_{21}$
is the Einstein A coefficient of $2.68\times10^{-7}$\,s$^{-1}$.
We find a total \Htwo\ molecular gas mass for both components of
$3.3\times10^{10}$\,M$_{\odot}$. We note that this value depends on the
adopted $Q_{21}$: for the highest $Q_{21}=1.0$, the $M($\Htwo) drops to
$2.2\times10^{10}$\,M$_{\odot}$, while for the lowest $Q_{21}=0.15$,
the $M($\Htwo) is $1.5\times10^{11}$\,M$_{\odot}$ The median value is
close to $M(\text{\Htwo})=6\times 10^{10}$\,M$_{\odot}$ derived from
CO(1-0) \citep{emonts13}. 
This total mass of diffuse molecular gas as probed by
the \CI\ emission is clearly higher than the warm ($T>300$\,K) molecular gas mass
of order $10^7$ to $10^9$\,M$_{\odot}$ \citep{ogle12} which suggest an order of magnitude ratio
of warm-to-cold molecular gas between $0.1-10$\%.  \cite{ogle12} suggest that
the gas is heated by the radio jet from the AGN interacting with the ambient
ISM of the \HzRG.  Furthermore, \cite{nesvadba06} find a mass of
ionised gas of $\sim3\times10^9$\,M$_{\odot}$. The gas masses of
various components of the ISM of the \HzRG\ are up to one order of 
magnitude lower than the stellar mass of $<1.8\times10^{12}$\,M$_{\odot}$ 
\citep{debreuck10}, suggesting that this is already a quite evolved galaxy.

For \compII, the \Htwo\ gas mass of the broad component is
$1.6\times10^{10}$\,M$_{\odot}$. We use this to constrain the size of
this region by knowing that $M(\text{\Htwo})<M_{\text{dyn}}=R\cdot v^2/G$,
and the velocity from the FWHM (see Table~\ref{table:lines}) we find that
$R<60$\,pc. Though uncertain, this region has a size comparable
to those of the obscuring torus and the inner narrow line regions of QSOs,
thus suggesting the presence of an additional radio quiet AGN, merging
with the radio loud AGN in \compI\ \citep{emonts15}. Applying the same
methods on \compI, which has a \Htwo\ gas mass of $1.7\times10^{10}$,
puts an upper limit of 1\,kpc on the size of the emitting region.

To determine the dust mass, we need to integrate over the full SED of the
dust emission. As we only have a single observation at \dustcont, we fix
the shape of the SED by assuming a greybody characterised by $\beta=1.5$
and $T_d=40$\,K. Following the approach of \cite{debreuck03}, using,

\begin{align}
M_d=\frac{S_{\text{obs}}D_L^2}{(1+z)\kappa_d(\nu_{\text{rest}})B(\nu_{\text{rest}},T_\text{d})},
\end{align}

where $S_{\text{obs}}$ is the flux density at the observed frequency,
$D_L$ is the luminosity distance, $\kappa_d(\nu_{\text{rest}})$ is the
rest-frequency mass absorption coefficient, which we here calculate by
scaling the 850\,$\mu$m absorption coefficient of 0.076\,m$^2$\,kg$^{-1}$
from \cite{stevens03} to \dustcont\, and $B(\nu_{\text{rest}},T_\text{d})$
is the the Planck function for isothermal dust grain emission at a
dust temperature $T_\text{d}$ \citep{debreuck03}, we derive a dust
mass of $M_d=6.1\times10^{8}$\,M$_{\odot}$ for the \HzRG. This is
marginally higher than what was estimated by \citealt{stevens03} of
$4.6\times10^{8}$\,M$_{\odot}$ using the 850\,$\mu$m flux peak flux.

Using the same approach and assumed dust temperature, we find the dust
masses of the two companions \#8 and \#10 (see Table~\ref{table:mass}).

\begin{table}      
\centering          
\begin{tabular}{l c c c}
\hline\hline       
Component                     & $M_d$             & L$_{\text{FIR}}$ & SFR \\
                                        &  [M$_{\odot}$] & [L$_{\odot}$]      &  [M$_{\odot}$\,yr$^{-1}$]        \\ 
\hline                    
\HzRG                             & $<6.1\times10^{8}$ & $<4.4\times10^{12}$ & $<440$ \\ 
\hline
companion \#8                & $<8.5\times10^{7}$ & $<6.2\times10^{11}$  & $<62$ \\ 
companion \#10              & $<6.5\times10^{7}$ & $<4.7\times10^{11}$  & $<47$ \\ 
\hline
Sum                            &    ---                        &  $<5.4\times10^{12}$ & $<549$ \\
\hline                    
Water (West)               & $<7.0\times10^{7}$ & $<5.7\times10^{11}$ & $<57$ \\ 
Water (East)                & $<7.0\times10^{7}$ & $<5.7\times10^{11}$ & $<57$ \\ 
\hline
\end{tabular}
\caption{{\small The estimated dust masses, FIR luminosities and
SFRs for the \HzRG\ and companions using the measured and listed in
Table~\ref{table:cont} and equation 1-5. The dust masses, FIR luminosities
and SFRs are all upper limits, as the \HzRG\ and companion \#10 have
contribution from synchrotron emission and companion \#8 is a tentative
detection.}}
\label{table:mass}
\end{table}

\subsection{Star-formation rate}

Using {\it Spitzer} and {\it Herschel} PACS+SPIRE photometry,
\citet{seymour12} showed that the strong AGN and vigorous star-formation
contribute with roughly equal shares to the IR luminosity in the \Spider.
While the AGN component dominates at the high frequency end of the
thermal dust emission, the star-formation is expected to contribute
$>90$\% of the flux at the observed (comparatively low) frequencies
of our ALMA data.  However, the flux spectral index of the \HzRG\ and
companion \#10, implies that the flux at these wavelength are likely
to be contaminated by synchrotron emission.  We can therefore use our
continuum flux to determine upper limits on the SFR in the host galaxy and the companions.

We estimate the SFR of the \HzRG\ and companions by fixing the dust
temperature and emissivity index (as for the dust masses), and calculating
$L_{\text{FIR}}$ using:

\begin{align}
L_{\text{FIR}}= 4\pi M_d\int^{\infty}_0\kappa_d(\nu_{\text{rest}})B(\nu_{\text{rest}},T_{\text{d}})d\nu,
\end{align}
which yields 
\begin{align}
L_{\text{FIR}}=4 \pi \Gamma[\beta+4] \zeta [\beta+4] D_L^2x^{-(\beta+4)}(e^x-1)S_{\text{obs}}\nu_{\text{obs}},
\end{align}
where $\Gamma$ and $\zeta$ are the Gamma and Riemann $\zeta$ functions
and $x=h\nu_{\text{rest}}/kT_{\text{d}}$ \citep{debreuck03}.
Again assuming $\beta=1.5$ and $T_{\text{d}}=40$\,K, we find
the $L_{\text{FIR}}<5.1\times10^{12}$\,L$_{\odot}$ for the
\HzRG, and $<7.2\times10^{11}$\,L$_{\odot}$ for companion \#8
and $<5.4\times10^{11}$\,L$_{\odot}$ for companion \#10 (see
Table~\ref{table:mass}).

Using,

\begin{align}
\text{SFR}=\delta_{\text{MF}}\delta_{\text{SB}}(L_{\text{FIR}}/10^{10}\,L_{\odot})\,M_{\odot}\,\text{yr}^{-1},
\end{align}
\noindent
where $\delta_{\text{MF}}$ is the stellar mass function and
$\delta_{\text{SB}}$ is the fraction of FIR emission heated by the
starburst, which we assume to be unity at these long wavelengths,
the $L_{\text{FIR}}$ can be converted to the SFR.  We assume
$\delta_{\text{MF}}=1$ and derive the SFRs of $<440$\,M$_{\odot}$ yr$^{-1}$
for the \HzRG, $<62$\,M$_{\odot}$ yr$^{-1}$ for companion \#8 and
$<47$\,M$_{\odot}$ yr$^{-1}$ for companion \#10 (see Table~\ref{table:mass}).
The sum of the estimated FIR luminosity of the \HzRG\ and the two
companion sources is higher than the total starburst-heated IR luminosity
of $5.6\times 10^{12}$L$_{\rm \odot}$ of \citealt{drouart14}, less than
the $7.9\times 10^{12}$L$_{\rm \odot}$ of \citet{seymour12}.  We warn
that our approximation of the SED with a single grey body function with
an assumed emissivity index is rather crude; moreover the \dustcont\
emission from the \HzRG\ and companion \#10 may also contain a
contribution from synchrotron emission, which would over-estimate the SFR.

\subsubsection{SFR of the \Spider\ and its companions}

\cite{kuiper11} found a bi-model velocity distribution of
galaxies close in projection to the \Spider\ which suggests two groups
(or regions) of galaxies flowing into the circum-galactic environment
at high velocity ($\sim400$ to 1600\,km/s).  Comparing the velocity and
spatial distribution of the galaxies with that found in massive halos
at $z\sim2$ within the Millennium Simulation, they conclude that the
\Spider\ is best described by the merger of two galaxy groups or flows of
galaxies into the over-dense environment of the powerful \HzRG.

The SFRs we determine here for the \HzRG\ and companion \#10 and \#8
are $\sim10 - 50$ times higher than determined by \cite{hatch09} using
optical \textit{HST} imaging. However, the continuum emission in both the
\HzRG\ and companion \#10 are contaminated by synchrotron emission.
For the \HzRG\ there are two components and only one of them, the
AGN core, is contaminated up to $\sim$50\% (see \S~\ref{res:dust}). The fact that the continuum emission is extended in the direction connecting these two components (see Fig.~\ref{fig:overview}) suggests that both must have roughly similar \dustcont\ flux densities. The synchrotron contamination for the combined \Spider\ is therefore estimated to be no more than $\sim$25\%.
For companion \#10, all of the emission could be due to
synchrotron (see \S~\ref{res:dust}). Observations of the synchrotron spectrum around $\sim$ 100\,GHz are needed to determine if it remains straight out to \dustcont, or if there is a down-turn due to energy losses of the most energetic electrons. In the latter case, even companion \#10 could still be dominated by star-formation. In this extreme case, then its SFR and that of companion
\#8 would suggest that (over the scale of our ALMA data) about $\sim20$\% of the star formation is taking place outside the \HzRG\
in companion sources. We note that companion \#8 is a marginal $3\sigma$ detection and that the two
additional companions (\#12 and \#17) show hints of \dustcont\ emission
at $\sim1.5\sigma$ significance.  To confirm if these hints of emission
are real requires deeper \dustcont\ observations. For the remaining sources in the surroundings of the \Spider, 
the sensitivity of our data is too low to detect them, meaning that their SFR is $\la$
50\,M$_{\sun}$\,yr$^{-1}$. 

\section{Conclusions}\label{sec:con}

We have observed the \HzRG\ \MRC\ (the \Spider) with ALMA Band 6 in
Cycle 1 for 49\,min on-source.

- We detect strong \dustcont\ continuum emission at the position of the
\HzRG, weaker \dustcont\ continuum emission for a companion to the west
and tentative emission to the north.  Companion \#10 of \citet{kuiper11}
at $z=2.1446$ is the brightest of the two companions and is detected in
\CO, while \CI\ is shifted out of the observed band. The other companion
is too weak to be detected in line emission.

- The flat spectral indexes of the \HzRG\ and companion \#10 suggest
a contribution to the \dustcont\ continuum emission from synchrotron
emission.  This means the FIR luminosities and SFRs of these two
components are upper limits.

- We detect strong \CI\ emission in two components: one at the position
of the \HzRG, and one with a small offset of 0\farcs5 to the
south-east from the central position. The \CI\ emitting components show
significantly different velocity profiles: \compI\ shows a Lorentzian
profile, and \compII\ is a double Gaussian with velocity widths of 250
and 1100\,km/s.

- Using the molecular gas mass estimate of the broad \CI\ of \compII,
we put an upper limit on the size of the emitting region of 60\,pc. This
size suggests the possible existence of an additional radio quiet AGN,
merging with the radio loud AGN in \compI.

- The \CO/\CI\ line ratio for the \Spider\ is lower than
in a comparison sample.  This ratio can be due to a relatively low
excitation molecular gas reservoir and/or a strong cosmic ray field which
preferentially destroys CO.  As the comparison sample is composed of
only lensed sources from the literature, differential lensing may also
have an influence, where the dense \CO\ gas in the comparison sample is
amplified more than the more diffuse \CI\ gas.

- We detect $\sim4\sigma$ \water\ emission lines at two positions: to the
west at knot B4 (at the bend) in the radio jet and to the east, upstream
of knot A.  No \dustcont\ continuum emission is detected at the position
of the \water\ emission lines, meaning that heating from IR emission
cannot be the source of the \water\ emission.  The positions
(upstream of the two knots in the radio jet), the lack of \dustcont\
continuum emission at these positions and the small offset of the line
centres, are consistent with heating by slow shocks
in dense molecular gas. The relatively bright \water\ emission indicates strong
dissipation of the energy of the expanding jets in the halo of \MRC.

Multiple studies have established that the \Spider\ is a system of merging
galaxies, and the likely progenitor of a cD galaxy. Our water
detections upstream of the main radio hotspots show that the radio jets
can cause the gas in the ISM and inter-cluster gas to cool, increasing
the fraction of extended molecular gas (Emonts et al. 2016, in prep.).
The formation of this molecular reservoir likely supports the formation
of the young stars observed in the halo of \MRC.  The much brighter than
expected \CI\ emission indicates the power of this line to study the
detailed distribution of the molecular gas reservoir in this intriguing
source, and possibly other \HzRG s.

\begin{acknowledgements}

This paper makes use of the following ALMA data:
ADS/JAO.ALMA\#2012.1.01087.S. ALMA is a partnership of ESO (representing
its member states), NSF (USA) and NINS (Japan), together with NRC
(Canada), NSC and ASIAA (Taiwan), and KASI (Republic of Korea), in
cooperation with the Republic of Chile. The Joint ALMA Observatory is
operated by ESO, AUI/NRAO and NAOJ.  
BG acknowledges support from the ERC Advanced Investigator programme DUSTYGAL 321334.
BE acknowledges funding through the European Union FP7-PEOPLE-2013-IEF grand 62435.

\end{acknowledgements}


\bibliographystyle{aa}
\bibliography{bibtex/HzRG_bib}

\begin{thebibliography}{54}
\expandafter\ifx\csname natexlab\endcsname\relax\def\natexlab#1{#1}\fi

\bibitem[{{Alaghband-Zadeh} {et~al.}(2013){Alaghband-Zadeh}, {Chapman},
  {Swinbank}, {Smail}, {Danielson}, {Decarli}, {Ivison}, {Meijerink}, {Weiss},
  \& {van der Werf}}]{alaghband-zadeh13}
{Alaghband-Zadeh}, S., {Chapman}, S.~C., {Swinbank}, A.~M., {et~al.} 2013,
  \mnras, 435, 1493

\bibitem[{{Alloin} {et~al.}(1997){Alloin}, {Guilloteau}, {Barvainis},
  {Antonucci}, \& {Tacconi}}]{alloin97}
{Alloin}, D., {Guilloteau}, S., {Barvainis}, R., {Antonucci}, R., \& {Tacconi},
  L. 1997, \aap, 321, 24

\bibitem[{{Appleton} {et~al.}(2013){Appleton}, {Guillard}, {Boulanger},
  {Cluver}, {Ogle}, {Falgarone}, {Pineau des For{\^e}ts}, {O'Sullivan}, {Duc},
  {Gallagher}, {Gao}, {Jarrett}, {Konstantopoulos}, {Lisenfeld}, {Lord}, {Lu},
  {Peterson}, {Struck}, {Sturm}, {Tuffs}, {Valchanov}, {van der Werf}, \&
  {Xu}}]{appleton13}
{Appleton}, P.~N., {Guillard}, P., {Boulanger}, F., {et~al.} 2013, \apj, 777,
  66

\bibitem[{{Bicknell} {et~al.}(1998){Bicknell}, {Dopita}, {Tsvetanov}, \&
  {Sutherland}}]{bicknell98}
{Bicknell}, G.~V., {Dopita}, M.~A., {Tsvetanov}, Z.~I., \& {Sutherland}, R.~S.
  1998, \apj, 495, 680

\bibitem[{{Bisbas} {et~al.}(2015){Bisbas}, {Papadopoulos}, \&
  {Viti}}]{bisbas15}
{Bisbas}, T.~G., {Papadopoulos}, P.~P., \& {Viti}, S. 2015, \apj, 803, 37

\bibitem[{{Carilli} {et~al.}(2002){Carilli}, {Harris}, {Pentericci},
  {R{\"o}ttgering}, {Miley}, {Kurk}, \& {van Breugel}}]{carilli02}
{Carilli}, C.~L., {Harris}, D.~E., {Pentericci}, L., {et~al.} 2002, \apj, 567,
  781

\bibitem[{{Carilli} {et~al.}(1997){Carilli}, {R{\"o}ttgering}, {van Ojik},
  {Miley}, {Breugel}, \& {W.~J.~M.~van}}]{carilli97}
{Carilli}, C.~L., {R{\"o}ttgering}, H.~J.~A., {van Ojik}, R., {et~al.} 1997,
  \apjs, 109, 1

\bibitem[{{Cox} {et~al.}(2011){Cox}, {Krips}, {Neri}, {Omont}, {G{\"u}sten},
  {Menten}, {Wyrowski}, {Wei{\ss}}, {Beelen}, {Gurwell}, {Dannerbauer},
  {Ivison}, {Negrello}, {Aretxaga}, {Hughes}, {Auld}, {Baes}, {Blundell},
  {Buttiglione}, {Cava}, {Cooray}, {Dariush}, {Dunne}, {Dye}, {Eales},
  {Frayer}, {Fritz}, {Gavazzi}, {Hopwood}, {Ibar}, {Jarvis}, {Maddox},
  {Micha{\l}owski}, {Pascale}, {Pohlen}, {Rigby}, {Smith}, {Swinbank}, {Temi},
  {Valtchanov}, {van der Werf}, \& {de Zotti}}]{cox11}
{Cox}, P., {Krips}, M., {Neri}, R., {et~al.} 2011, \apj, 740, 63

\bibitem[{{Danielson} {et~al.}(2011){Danielson}, {Swinbank}, {Smail}, {Cox},
  {Edge}, {Weiss}, {Harris}, {Baker}, {De Breuck}, {Geach}, {Ivison}, {Krips},
  {Lundgren}, {Longmore}, {Neri}, \& {Flaquer}}]{danielson11}
{Danielson}, A.~L.~R., {Swinbank}, A.~M., {Smail}, I., {et~al.} 2011, \mnras,
  410, 1687

\bibitem[{{Dannerbauer} {et~al.}(2014){Dannerbauer}, {Kurk}, {De Breuck},
  {Wylezalek}, {Santos}, {Koyama}, {Seymour}, {Tanaka}, {Hatch}, {Altieri},
  {Coia}, {Galametz}, {Kodama}, {Miley}, {R{\"o}ttgering}, {Sanchez-Portal},
  {Valtchanov}, {Venemans}, \& {Ziegler}}]{dannerbauer14}
{Dannerbauer}, H., {Kurk}, J.~D., {De Breuck}, C., {et~al.} 2014, \aap, 570,
  A55

\bibitem[{{De Breuck} {et~al.}(2003){De Breuck}, {Neri}, {Morganti}, {Omont},
  {Rocca-Volmerange}, {Stern}, {Reuland}, {van Breugel}, {R{\"o}ttgering},
  {Stanford}, {Spinrad}, {Vigotti}, \& {Wright}}]{debreuck03}
{De Breuck}, C., {Neri}, R., {Morganti}, R., {et~al.} 2003, \aap, 401, 911

\bibitem[{{De Breuck} {et~al.}(2010){De Breuck}, {Seymour}, {Stern}, {Willner},
  {Eisenhardt}, {Fazio}, {Galametz}, {Lacy}, {Rettura}, {Rocca-Volmerange}, \&
  {Vernet}}]{debreuck10}
{De Breuck}, C., {Seymour}, N., {Stern}, D., {et~al.} 2010, \apj, 725, 36

\bibitem[{{Downes} \& {Solomon}(2003)}]{downes03}
{Downes}, D. \& {Solomon}, P.~M. 2003, \apj, 582, 37

\bibitem[{{Drouart} {et~al.}(2014){Drouart}, {De Breuck}, {Vernet}, {Seymour},
  {Lehnert}, {Barthel}, {Bauer}, {Ibar}, {Galametz}, {Haas}, {Hatch},
  {Mullaney}, {Nesvadba}, {Rocca-Volmerange}, {R{\"o}ttgering}, {Stern}, \&
  {Wylezalek}}]{drouart14}
{Drouart}, G., {De Breuck}, C., {Vernet}, J., {et~al.} 2014, \aap, 566, A53

\bibitem[{{Emonts} {et~al.}(2015){Emonts}, {De Breuck}, {Lehnert}, {Vernet},
  {Gullberg}, {Villar-Mart{\'{\i}}n}, {Nesvadba}, {Drouart}, {Ivison},
  {Seymour}, {Wylezalek}, \& {Barthel}}]{emonts15}
{Emonts}, B.~H.~C., {De Breuck}, C., {Lehnert}, M.~D., {et~al.} 2015, ArXiv
  e-prints

\bibitem[{{Emonts} {et~al.}(2013){Emonts}, {Feain}, {R{\"o}ttgering}, {Miley},
  {Seymour}, {Norris}, {Carilli}, {Villar-Mart{\'{\i}}n}, {Mao}, {Sadler},
  {Ekers}, {van Moorsel}, {Ivison}, {Pentericci}, {Tadhunter}, \&
  {Saikia}}]{emonts13}
{Emonts}, B.~H.~C., {Feain}, I., {R{\"o}ttgering}, H.~J.~A., {et~al.} 2013,
  \mnras, 430, 3465

\bibitem[{{Emonts} {et~al.}(2014){Emonts}, {Norris}, {Feain}, {Mao}, {Ekers},
  {Miley}, {Seymour}, {R{\"o}ttgering}, {Villar-Mart{\'{\i}}n}, {Sadler},
  {Carilli}, {Mahony}, {de Breuck}, {Stroe}, {Pentericci}, {van Moorsel},
  {Drouart}, {Ivison}, {Greve}, {Humphrey}, {Wylezalek}, \&
  {Tadhunter}}]{emonts14}
{Emonts}, B.~H.~C., {Norris}, R.~P., {Feain}, I., {et~al.} 2014, \mnras, 438,
  2898

\bibitem[{{Ferkinhoff} {et~al.}(2014){Ferkinhoff}, {Brisbin}, {Parshley},
  {Nikola}, {Stacey}, {Schoenwald}, {Higdon}, {Higdon}, {Verma}, {Riechers},
  {Hailey-Dunsheath}, {Menten}, {G{\"u}sten}, {Wei{\ss}}, {Irwin}, {Cho},
  {Niemack}, {Halpern}, {Amiri}, {Hasselfield}, {Wiebe}, {Ade}, \&
  {Tucker}}]{ferkinhoff14}
{Ferkinhoff}, C., {Brisbin}, D., {Parshley}, S., {et~al.} 2014, \apj, 780, 142

\bibitem[{{Flower} \& {Pineau Des For{\^e}ts}(2010)}]{flower10}
{Flower}, D.~R. \& {Pineau Des For{\^e}ts}, G. 2010, \mnras, 406, 1745

\bibitem[{{Glover} {et~al.}(2015){Glover}, {Clark}, {Micic}, \&
  {Molina}}]{glover15}
{Glover}, S.~C.~O., {Clark}, P.~C., {Micic}, M., \& {Molina}, F. 2015, \mnras,
  448, 1607

\bibitem[{{Goicoechea} {et~al.}(2015){Goicoechea}, {Chavarr{\'{\i}}a},
  {Cernicharo}, {Neufeld}, {Vavrek}, {Bergin}, {Cuadrado}, {Encrenaz},
  {Etxaluze}, {Melnick}, \& {Polehampton}}]{goicoechea15}
{Goicoechea}, J.~R., {Chavarr{\'{\i}}a}, L., {Cernicharo}, J., {et~al.} 2015,
  \apj, 799, 102

\bibitem[{{Greve} {et~al.}(2006){Greve}, {Ivison}, \& {Stevens}}]{greve06}
{Greve}, T.~R., {Ivison}, R.~J., \& {Stevens}, J.~A. 2006, Astronomische
  Nachrichten, 327, 208

\bibitem[{{Guillard} {et~al.}(2010){Guillard}, {Boulanger}, {Cluver},
  {Appleton}, {Pineau Des For{\^e}ts}, \& {Ogle}}]{guillard10}
{Guillard}, P., {Boulanger}, F., {Cluver}, M.~E., {et~al.} 2010, \aap, 518, A59

\bibitem[{{Guillard} {et~al.}(2009){Guillard}, {Boulanger}, {Pineau Des
  For{\^e}ts}, \& {Appleton}}]{guillard09}
{Guillard}, P., {Boulanger}, F., {Pineau Des For{\^e}ts}, G., \& {Appleton},
  P.~N. 2009, \aap, 502, 515

\bibitem[{{Guillard} {et~al.}(2012){Guillard}, {Boulanger}, {Pineau des
  For{\^e}ts}, {Falgarone}, {Gusdorf}, {Cluver}, {Appleton}, {Lisenfeld},
  {Duc}, {Ogle}, \& {Xu}}]{guillard12}
{Guillard}, P., {Boulanger}, F., {Pineau des For{\^e}ts}, G., {et~al.} 2012,
  \apj, 749, 158

\bibitem[{{Hatch} {et~al.}(2009){Hatch}, {Overzier}, {Kurk}, {Miley},
  {R{\"o}ttgering}, \& {Zirm}}]{hatch09}
{Hatch}, N.~A., {Overzier}, R.~A., {Kurk}, J.~D., {et~al.} 2009, \mnras, 395,
  114

\bibitem[{{Hatch} {et~al.}(2008){Hatch}, {Overzier}, {R{\"o}ttgering}, {Kurk},
  \& {Miley}}]{hatch08}
{Hatch}, N.~A., {Overzier}, R.~A., {R{\"o}ttgering}, H.~J.~A., {Kurk}, J.~D.,
  \& {Miley}, G.~K. 2008, \mnras, 383, 931

\bibitem[{{Humphrey} {et~al.}(2008){Humphrey}, {Villar-Mart{\'{\i}}n},
  {Vernet}, {Fosbury}, {di Serego Alighieri}, \& {Binette}}]{humphrey08}
{Humphrey}, A., {Villar-Mart{\'{\i}}n}, M., {Vernet}, J., {et~al.} 2008,
  \mnras, 383, 11

\bibitem[{{Kneib} {et~al.}(2005){Kneib}, {Neri}, {Smail}, {Blain}, {Sheth},
  {van der Werf}, \& {Knudsen}}]{kneib05}
{Kneib}, J.-P., {Neri}, R., {Smail}, I., {et~al.} 2005, \aap, 434, 819

\bibitem[{{Kuiper} {et~al.}(2011){Kuiper}, {Hatch}, {Miley}, {Nesvadba},
  {R{\"o}ttgering}, {Kurk}, {Lehnert}, {Overzier}, {Pentericci}, {Schaye}, \&
  {Venemans}}]{kuiper11}
{Kuiper}, E., {Hatch}, N.~A., {Miley}, G.~K., {et~al.} 2011, \mnras, 415, 2245

\bibitem[{{Kurk} {et~al.}(2004{\natexlab{a}}){Kurk}, {Pentericci}, {Overzier},
  {R{\"o}ttgering}, \& {Miley}}]{kurk04b}
{Kurk}, J.~D., {Pentericci}, L., {Overzier}, R.~A., {R{\"o}ttgering}, H.~J.~A.,
  \& {Miley}, G.~K. 2004{\natexlab{a}}, \aap, 428, 817

\bibitem[{{Kurk} {et~al.}(2004{\natexlab{b}}){Kurk}, {Pentericci},
  {R{\"o}ttgering}, \& {Miley}}]{kurk04a}
{Kurk}, J.~D., {Pentericci}, L., {R{\"o}ttgering}, H.~J.~A., \& {Miley}, G.~K.
  2004{\natexlab{b}}, \aap, 428, 793

\bibitem[{{Lestrade} {et~al.}(2010){Lestrade}, {Combes}, {Salom{\'e}}, {Omont},
  {Bertoldi}, {Andr{\'e}}, \& {Schneider}}]{lestrade10}
{Lestrade}, J.-F., {Combes}, F., {Salom{\'e}}, P., {et~al.} 2010, \aap, 522, L4

\bibitem[{{Lonsdale} \& {Barthel}(1986)}]{lonsdale86}
{Lonsdale}, C.~J. \& {Barthel}, P.~D. 1986, \aj, 92, 12

\bibitem[{{Lupu} {et~al.}(2012){Lupu}, {Scott}, {Aguirre}, {Aretxaga}, {Auld},
  {Barton}, {Beelen}, {Bertoldi}, {Bock}, {Bonfield}, {Bradford},
  {Buttiglione}, {Cava}, {Clements}, {Cooke}, {Cooray}, {Dannerbauer},
  {Dariush}, {De Zotti}, {Dunne}, {Dye}, {Eales}, {Frayer}, {Fritz}, {Glenn},
  {Hughes}, {Ibar}, {Ivison}, {Jarvis}, {Kamenetzky}, {Kim}, {Lagache},
  {Leeuw}, {Maddox}, {Maloney}, {Matsuhara}, {Murphy}, {Naylor}, {Negrello},
  {Nguyen}, {Omont}, {Pascale}, {Pohlen}, {Rigby}, {Rodighiero}, {Serjeant},
  {Smith}, {Temi}, {Thompson}, {Valtchanov}, {Verma}, {Vieira}, \&
  {Zmuidzinas}}]{lupu12}
{Lupu}, R.~E., {Scott}, K.~S., {Aguirre}, J.~E., {et~al.} 2012, \apj, 757, 135

\bibitem[{{Miley} {et~al.}(2006){Miley}, {Overzier}, {Zirm}, {Ford}, {Kurk},
  {Pentericci}, {Blakeslee}, {Franx}, {Illingworth}, {Postman}, {Rosati},
  {R{\"o}ttgering}, {Venemans}, \& {Helder}}]{miley06}
{Miley}, G.~K., {Overzier}, R.~A., {Zirm}, A.~W., {et~al.} 2006, \apjl, 650,
  L29

\bibitem[{{Nesvadba} {et~al.}(2006){Nesvadba}, {Lehnert}, {Eisenhauer},
  {Gilbert}, {Tecza}, \& {Abuter}}]{nesvadba06}
{Nesvadba}, N.~P.~H., {Lehnert}, M.~D., {Eisenhauer}, F., {et~al.} 2006, \apj,
  650, 693

\bibitem[{{Ogle} {et~al.}(2012){Ogle}, {Davies}, {Appleton}, {Bertincourt},
  {Seymour}, \& {Helou}}]{ogle12}
{Ogle}, P., {Davies}, J.~E., {Appleton}, P.~N., {et~al.} 2012, \apj, 751, 13

\bibitem[{{Omont} {et~al.}(2013){Omont}, {Yang}, {Cox}, {Neri}, {Beelen},
  {Bussmann}, {Gavazzi}, {van der Werf}, {Riechers}, {Downes}, {Krips}, {Dye},
  {Ivison}, {Vieira}, {Wei{\ss}}, {Aguirre}, {Baes}, {Baker}, {Bertoldi},
  {Cooray}, {Dannerbauer}, {De Zotti}, {Eales}, {Fu}, {Gao}, {Gu{\'e}lin},
  {Harris}, {Jarvis}, {Lehnert}, {Leeuw}, {Lupu}, {Menten}, {Micha{\l}owski},
  {Negrello}, {Serjeant}, {Temi}, {Auld}, {Dariush}, {Dunne}, {Fritz},
  {Hopwood}, {Hoyos}, {Ibar}, {Maddox}, {Smith}, {Valiante}, {Bock},
  {Bradford}, {Glenn}, \& {Scott}}]{omont13}
{Omont}, A., {Yang}, C., {Cox}, P., {et~al.} 2013, \aap, 551, A115

\bibitem[{{Papadopoulos} {et~al.}(2004){Papadopoulos}, {Thi}, \&
  {Viti}}]{papadopoulos04}
{Papadopoulos}, P.~P., {Thi}, W.-F., \& {Viti}, S. 2004, \mnras, 351, 147

\bibitem[{{Pentericci} {et~al.}(2002){Pentericci}, {Kurk}, {Carilli}, {Harris},
  {Miley}, \& {R{\"o}ttgering}}]{pentericci02}
{Pentericci}, L., {Kurk}, J.~D., {Carilli}, C.~L., {et~al.} 2002, \aap, 396,
  109

\bibitem[{{Pentericci} {et~al.}(2000){Pentericci}, {Kurk}, {R{\"o}ttgering},
  {Miley}, {van Breugel}, {Carilli}, {Ford}, {Heckman}, {McCarthy}, \&
  {Moorwood}}]{pentericci00}
{Pentericci}, L., {Kurk}, J.~D., {R{\"o}ttgering}, H.~J.~A., {et~al.} 2000,
  \aap, 361, L25

\bibitem[{{Pentericci} {et~al.}(1997){Pentericci}, {R{\"o}ttgering}, {Miley},
  {Carilli}, \& {McCarthy}}]{pentericci97}
{Pentericci}, L., {R{\"o}ttgering}, H.~J.~A., {Miley}, G.~K., {Carilli}, C.~L.,
  \& {McCarthy}, P. 1997, \aap, 326, 580

\bibitem[{{Pentericci} {et~al.}(1998){Pentericci}, {R{\"o}ttgering}, {Miley},
  {Spinrad}, {McCarthy}, {van Breugel}, \& {Macchetto}}]{pentericci98}
{Pentericci}, L., {R{\"o}ttgering}, H.~J.~A., {Miley}, G.~K., {et~al.} 1998,
  \apj, 504, 139

\bibitem[{{Riechers} {et~al.}(2013){Riechers}, {Bradford}, {Clements},
  {Dowell}, {P{\'e}rez-Fournon}, {Ivison}, {Bridge}, {Conley}, {Fu}, {Vieira},
  {Wardlow}, {Calanog}, {Cooray}, {Hurley}, {Neri}, {Kamenetzky}, {Aguirre},
  {Altieri}, {Arumugam}, {Benford}, {B{\'e}thermin}, {Bock}, {Burgarella},
  {Cabrera-Lavers}, {Chapman}, {Cox}, {Dunlop}, {Earle}, {Farrah}, {Ferrero},
  {Franceschini}, {Gavazzi}, {Glenn}, {Solares}, {Gurwell}, {Halpern},
  {Hatziminaoglou}, {Hyde}, {Ibar}, {Kov{\'a}cs}, {Krips}, {Lupu}, {Maloney},
  {Martinez-Navajas}, {Matsuhara}, {Murphy}, {Naylor}, {Nguyen}, {Oliver},
  {Omont}, {Page}, {Petitpas}, {Rangwala}, {Roseboom}, {Scott}, {Smith},
  {Staguhn}, {Streblyanska}, {Thomson}, {Valtchanov}, {Viero}, {Wang},
  {Zemcov}, \& {Zmuidzinas}}]{riechers13}
{Riechers}, D.~A., {Bradford}, C.~M., {Clements}, D.~L., {et~al.} 2013, \nat,
  496, 329

\bibitem[{{R{\"o}ttgering} {et~al.}(1997){R{\"o}ttgering}, {van Ojik}, {Miley},
  {Chambers}, {van Breugel}, \& {de Koff}}]{roettgering97}
{R{\"o}ttgering}, H.~J.~A., {van Ojik}, R., {Miley}, G.~K., {et~al.} 1997,
  \aap, 326, 505

\bibitem[{{Salom{\'e}} {et~al.}(2011){Salom{\'e}}, {Combes}, {Revaz}, {Downes},
  {Edge}, \& {Fabian}}]{salome11}
{Salom{\'e}}, P., {Combes}, F., {Revaz}, Y., {et~al.} 2011, \aap, 531, A85

\bibitem[{{Serjeant}(2012)}]{serjeant12}
{Serjeant}, S. 2012, \mnras, 424, 2429

\bibitem[{{Seymour} {et~al.}(2012){Seymour}, {Altieri}, {De Breuck}, {Barthel},
  {Coia}, {Conversi}, {Dannerbauer}, {Dey}, {Dickinson}, {Drouart}, {Galametz},
  {Greve}, {Haas}, {Hatch}, {Ibar}, {Ivison}, {Jarvis}, {Kov{\'a}cs}, {Kurk},
  {Lehnert}, {Miley}, {Nesvadba}, {Rawlings}, {Rettura}, {R{\"o}ttgering},
  {Rocca-Volmerange}, {S{\'a}nchez-Portal}, {Santos}, {Stern}, {Stevens},
  {Valtchanov}, {Vernet}, \& {Wylezalek}}]{seymour12}
{Seymour}, N., {Altieri}, B., {De Breuck}, C., {et~al.} 2012, \apj, 755, 146

\bibitem[{{Stevens} {et~al.}(2003){Stevens}, {Ivison}, {Dunlop}, {Smail},
  {Percival}, {Hughes}, {R{\"o}ttgering}, {van Breugel}, \&
  {Reuland}}]{stevens03}
{Stevens}, J.~A., {Ivison}, R.~J., {Dunlop}, J.~S., {et~al.} 2003, \nat, 425,
  264

\bibitem[{{van der Werf} {et~al.}(2011){van der Werf}, {Berciano Alba},
  {Spaans}, {Loenen}, {Meijerink}, {Riechers}, {Cox}, {Wei{\ss}}, \&
  {Walter}}]{vanderwerf11}
{van der Werf}, P.~P., {Berciano Alba}, A., {Spaans}, M., {et~al.} 2011, \apjl,
  741, L38

\bibitem[{{Walter} {et~al.}(2011){Walter}, {Wei{\ss}}, {Downes}, {Decarli}, \&
  {Henkel}}]{walter11}
{Walter}, F., {Wei{\ss}}, A., {Downes}, D., {Decarli}, R., \& {Henkel}, C.
  2011, \apj, 730, 18

\bibitem[{{Wei{\ss}} {et~al.}(2003){Wei{\ss}}, {Henkel}, {Downes}, \&
  {Walter}}]{weiss03}
{Wei{\ss}}, A., {Henkel}, C., {Downes}, D., \& {Walter}, F. 2003, \aap, 409,
  L41

\bibitem[{{Yang} {et~al.}(2013){Yang}, {Gao}, {Omont}, {Liu}, {Isaak},
  {Downes}, {van der Werf}, \& {Lu}}]{yang13}
{Yang}, C., {Gao}, Y., {Omont}, A., {et~al.} 2013, \apjl, 771, L24

\end{thebibliography}

\end{document}